\documentclass[10pt,letterpaper,twocolumn,aps,showpacs,preprintnumbers,amsmath,amssymb,nofootinbib,prl]{revtex4-2}
\usepackage[utf8]{inputenc}
\usepackage{color}
\usepackage{amsmath}
\usepackage{siunitx}
\usepackage{amssymb}
\usepackage{graphicx}
\usepackage{xcolor}
\usepackage[unicode=true,pdfusetitle,
 bookmarks=true,bookmarksnumbered=false,bookmarksopen=false,
 breaklinks=false,pdfborder={0 0 0},pdfborderstyle={},backref=false,colorlinks=true]
 {hyperref}
\hypersetup{
 pdfborderstyle={},colorlinks,linkcolor=red,citecolor=blue}

\makeatletter

\usepackage{dcolumn}
\usepackage{mathrsfs}
\usepackage{bm}

\usepackage{siunitx}

\usepackage{ulem}

\usepackage{xr}
\usepackage{soul}

\makeatother

\begin{document}
\title{Atom-light hybrid interferometer for atomic sensing with quantum memory}

\author{Xingchang Wang,$^{1,2,*,\S}$ Xinyun Liang,$^{1,2,*}$ Dasen Yang,$^{1,2,*}$ Liang Dong,$^{1,2}$ Ying Zuo,$^{2}$ Jianmin Wang,$^{3}$ Linyu Chen,$^{1,2}$ Georgios A. Siviloglou,$^{4,5,1,2}$ Z. Y. Ou,$^{3,\ddagger}$ and J. F. Chen,$^{1,2,\dagger}$}

\affiliation{$^{1}$Department of Physics, Southern University of Science and Technology, Shenzhen 518055, China\\ 
$^{2}$International Shenzhen Quantum Academy, Shenzhen 518048, China\\ 
$^{3}$Department of Physics, City University of Hong Kong, 83 Tat Chee Avenue, Kowloon, Hong Kong, China\\ 
$^{4}$Department of Physics, University of Crete, 71003, Heraklion, Greece\\ 
$^{5}$Institute of Electronic Structure and Laser (IESL), FORTH, 71110, Heraklion, Greece}
\date{\today}

\begin{abstract}
Quantum memories feature a reversible conversion of optical fields into long-lived atomic spin waves, and are therefore ideal for operating as sensitive atomic sensors. However, up to now, atom-light interferometers have lacked an efficient approach to exploit their ultimate atomic sensing performance, since an extra optical delay line is required to compensate for the memory time. Here, we report a new protocol that records the photocurrent via heterodyne mixing with a stable local oscillator. The obtained complex quadrature amplitude that carries information imprinted on its phase by an external magnetic field, is successfully recovered from the interference patterns between the light and the atomic spin wave, without the stringent requirement of having them overlap in time. Our results reveal that the sensitivity scales favorably with the lifetime of the quantum memory. Our work may have important applications in building distributed quantum networks through quantum memory-assisted atom-light interferometers.
\end{abstract}
\maketitle

Atom-light interferometers utilize flying photons and atoms carrying atomic coherence as two distinct interferometric arms. In such systems, photons serve as carriers of quantum information capable of propagating over long distances, while localized atomic qubits act as efficient units for computation and sensing. The recent developments in atom-light hybrid interferometers have demonstrated that the transformation processes compatible with memory implement coherent conversion between photons and long-lived atomic coherence and reproduce first- and second-order interference between atoms and light~\cite{ECNU_hybrid2015,Wen2019PRL,Wang2022Quantum,Su2022Quantum}. Using the core principle common to both optical and atomic interferometry, namely interference between atoms and between light pulses, atomic sensors can be used to precisely probe diverse physical quantities from magnetic or electric fields and from gravitation to inertia ~\cite{Chen2010Heralded,Kuan2016Large,Gomez2020Bose,ZTLuNP_Cat_2025,Ma2025Adaptive}. The atomic arm of an atom-light hybrid interferometer is well qualified to implement sensing, provided that the phase difference between the atomic and the photonic states is measurable.

The stored atomic coherence is considered as an atomic spin wave converted from an optical wave, and the lifetime of this coherence ranges from sub-microseconds to seconds for atomic ensembles~\cite{Zhao2009Millisecond,Zhao2009Longlived,Dudin2013Light,Yang2016Efficient,Cho2016Highly,Hsiao2018Highly}. Following phase sensing within the memory lifetime, the atomic spin wave needs to be combined with the other optical arm at the second beam splitter to regenerate interference. Critically, this architecture faces a fundamental challenge: the read-out signal from the stored atomic spin state is temporally separated from the photons emitted earlier. Only if these photons propagate through a long-distance optical delay path can they interfere with the atomic state. Although interference can in principle be restored using filtering and the aforementioned optical delay lines~\cite{Erdmann2000Restoring}, photon transmission losses severely degrade the interference visibility as the storage time increases. Consequently, effective encoding of phase information in existing implementations is constrained to the optical arm, whereas the atomic arm has yet to demonstrate efficient sensing. Consequently, due to lack of efficient protocol of local calibration method, the architecture of atom-light interferometer utilizing atom sensing remains an open question.

To avoid long optical delay lines, we propose to let the atom-light interferometer work as an unbalanced Mach-Zehnder interferometer (MZI), complemented by homodyne detection (HD) with a strong coherent local oscillator (LO) and the recording of the quadrature amplitude and phase of the optical field of the interferometer~\cite{Yuen1978Optical}. It has recently been demonstrated that the direct addition of the photocurrent appearing in HD could still lead to interference due to the addition of the amplitudes of the light fields, even though the path imbalance of the MZI is beyond coherence length of the input field~\cite{Zhang2025Optical,Tang2025Phasedependent}. By recording the quadrature amplitude instead of the intensity of the optical field, phase information is preserved in the electronic signals converted from photodetectors. In the past, similar schemes of MZIs with HD have been used to measure the phase shifts appearing in the dispersive regime of neutral atoms~\cite{Xiao1995Measurement,Zibrov1996Experimental,Aljunid2009Phase,Appel2008Quantum}. \par

\begin{figure}[htbp]
    \centering
    \includegraphics[width=1.00\linewidth]{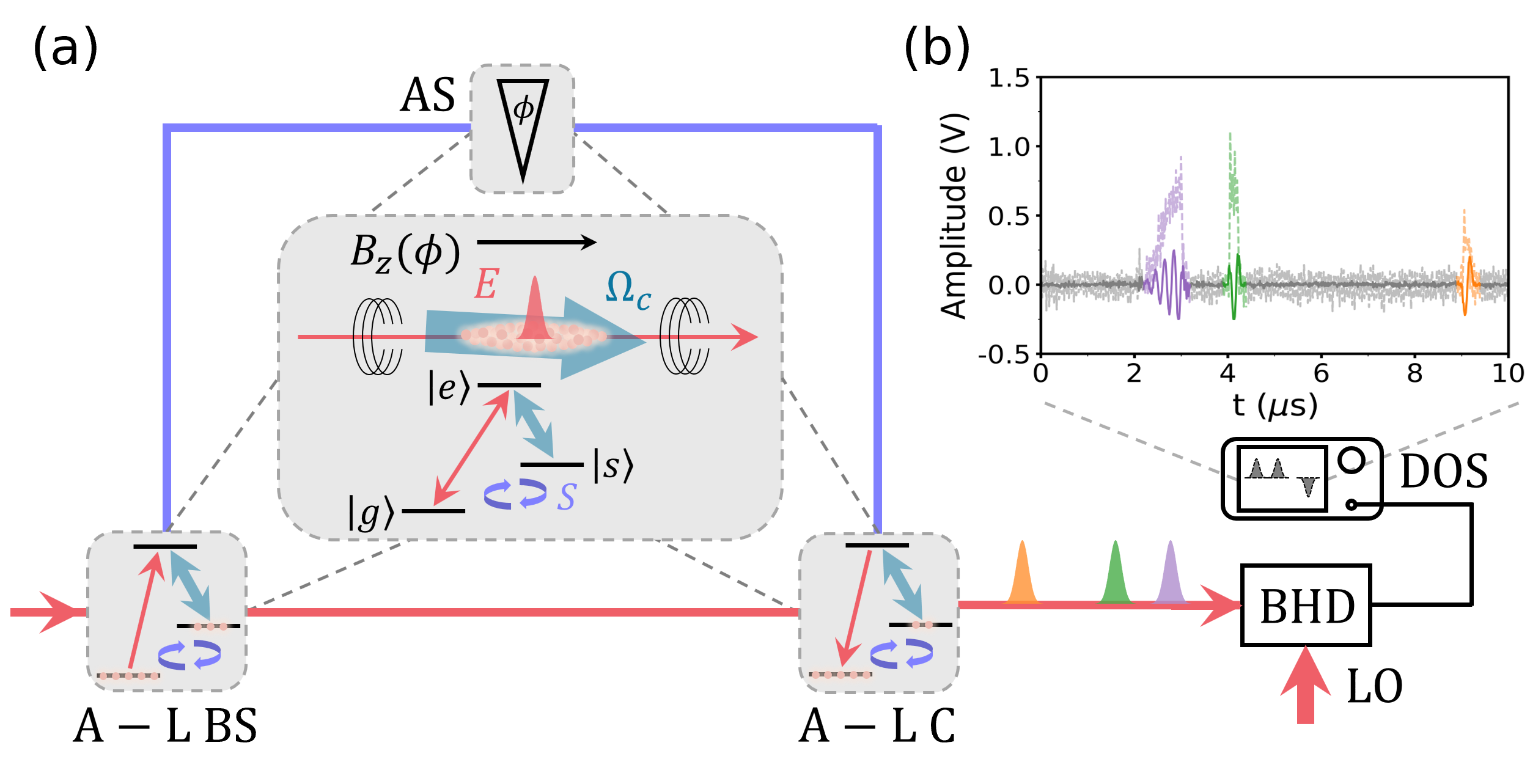}
    \caption{Schematic setup. (a) An elongated, laser-cooled atomic ensemble serves as an atom-light beam splitter (A-L BS), atomic sensor (AS), and coupler (A-L C). A strong control light denoted as $\Omega_c$ couples the signal field $E$ and atomic coherence denoted as $S$, which accumulates a phase $\phi$ in the presence of an external magnetic field $B_{z}$ in the $z$-direction. The balanced homodyne detector (BHD) with a strong local oscillator (LO) records the quadrature amplitude and phase of the light signal. (b) The raw electronic output of BHD on a digital oscilloscope (DOS) is shown as a solid line. The dashed line is pulse profile recorded by photomultiplier. The purple, green, and orange lines represent the transmitted signal and multiple readouts of the memory.}
    \label{fig:setup}
\end{figure}

In this Letter, we propose and demonstrate a new measurement protocol for atomic sensing based on an atom-light hybrid interferometer and amplitude measurement by HD. We utilize an elongated, laser-cooled atomic ensemble in a magneto-optical trap (MOT), in which a linear beam-splitter-type atom-light interaction is applied and photons are converted into atomic coherence with a phase controlled by an external magnetic field. The amplitude and phase of the transmitted and retrieved optical fields are recorded by beating with an LO in a balanced homodyne detector (BHD) as introduced in Ref.~\cite{Zhang2025Optical}. For the proof-of-principle, we show that the magnetic-field-controlled interference pattern is recovered by direct amplitude addition of the photocurrents without a second beam splitter and an optical delay line. The systematic errors and classical noise are well-controlled in our system, so that the precision of the magnetic field measurements is inversely proportional to the memory lifetime. The results demonstrate a practical protocol of local sensor calibration for atom-light interferometer. The atomic source with a long memory lifetime can enable distributed quantum sensing~\cite{doi:10.1126/science.1104149,XueshiNP2020,yuao2021NPdistributed,Malia2022Distributed}. Some plausible schemes have been put forward based on quantum communication using entangled particles \cite{Zhuang_2021,Network_XuPRL_2024}, and quantum repeaters \cite{LongBaseLine2012,Stas2026Entanglementassisted,Wang2026Memoryassisted}. 

Similar to unbalanced MZIs, we consider the time-separated signal fields transmitted (T) and retrieved (R) from the atomic ensemble; both of them appear as the optical mode outputs with amplitude $E$ and frequency $\omega_{S}$ of the atom-light interferometer: $E_{T}a_{T}(t)e^{-i\omega_{S}t}$ and $E_{R}a_{R}(t-\Delta\tau)e^{-i\omega_{S}(t-\Delta\tau)}$. Here, $a_{j}(t)$, $(j=T,R)$ are the normalized pulse profiles, and $\Delta\tau$ is the storage time for the atomic spin mode of the atom-light interferometer. In particular, the output of the HD is $i_{HD}(t)\propto|\mathcal{E}|X(t)$, where $|\mathcal{E}|$ is the amplitude of the LO and $X(t)=Ea(t)e^{-i\delta\omega t}e^{i\delta\phi}+c. c.$ is the quadrature-phase amplitude of the signal field; $\delta\omega=\omega_{LO}-\omega_{S}$ and $\delta\phi=\phi_{LO}-\phi_{S}$ are the frequency and phase differences between the signal field and the LO, respectively. We measure from the HD the photocurrent signals of the balanced detector as $i_{T}(t)\propto |\mathcal{E}|X_{T}(t)$ and $i_{R}(t)\propto |\mathcal{E}|X_{R}(t)$. By adding them together, we have $i_{+}(t)=i_{T}(t)+i_{R}(t)\propto|\mathcal{E}|[X_{T}(t)+X_{R}(t)]$. The power of the photocurrent is proportional to the average of $i_{+}^{2}(t)$, which is (see \hyperref[appendix]{Appendix}) 
\begin{equation}\label{eq:iave}
    \begin{aligned}
        \langle i_{+}^{2}(t)\rangle
        \approx 2K|\mathcal{E}|^{2}(I_{T}+I_{R})[1+\mathcal{V}_{0}\cos{(\delta\omega\Delta\tau-\Delta\phi)}], 
    \end{aligned}
\end{equation}
where $K$ corresponds to the response of the homodyne detector, $\mathcal{V}_{0}=2\sqrt{I_{T}I_{R}}/(I_{T}+I_{R})$ is the interference visibility, $I_{j}=|E_{j}|^{2}, (j=T,R)$ are the light intensities, and $\Delta\phi=\delta\phi_{T}-\delta\phi_{R}$ is the accumulated phase during storage.

\begin{figure}[!ht]
    \centering
    \includegraphics[width=1.0\linewidth]{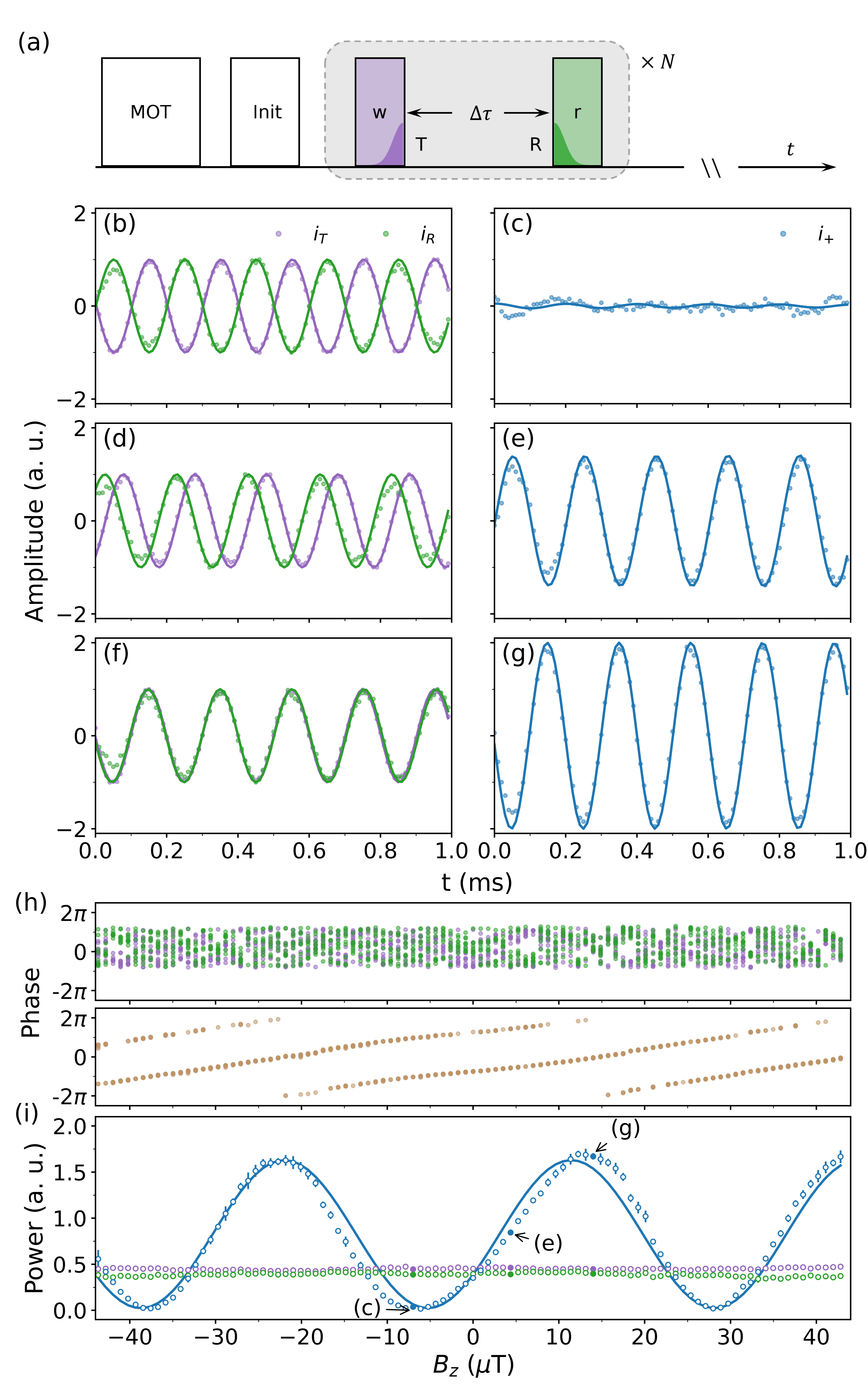}
    \caption{(a) The measurement sequence for directly comparing transmitted and retrieved pulses, in which the storage time is $\Delta\tau=\SI{5}{\mu s}$. (b)-(g) Beating patterns between the signal and local oscillator in the balanced detector with a frequency difference {$\delta\omega/2\pi=\SI{5}{kHz}$}; the purple and green dots are experimental data for the transmitted $i_T$ and retrieved signal $i_R$, and the solid curve represents a fit to the data. The phase differences for (b): $\pi$, (d): $\frac{\pi}{2}$, and (f): $0$. (c), (e), and (g) are the amplitudes of the addition in blue, corresponding to the filled points in (i). (h) The phase difference between LO and T is plotted as purple dots, and that between LO and R is plotted as green dots. The relative phase between T and R is plotted as yellow dots as a function of the scanned magnetic field. (i) Recovery of the interference fringes using the average power of the addition of the T and R beating patterns. The dots represent experimental data, and the solid line is the fitted curve. }
    \label{fig:fringes}
\end{figure}

The stored collective excitation in an ensemble of $N_a$ atoms can be written as $|\psi (t)\rangle=\hat{S} ^{\dagger} (t)|g_{1},\cdots,g_{j},\cdots,g_{N_{a}}\rangle$, where 
$\hat{S} (t) = (1/\sqrt{N_a})\sum_{j} \exp[-i\Delta\mathbf{k}\cdot \mathbf{r}_{j} (t)+i\Delta\omega_{gs}t]\left | g \right \rangle_j \left\langle s \right|$ is the collective atomic excitation operator of the spin wave. $\Delta\mathbf{k}=\mathbf{k}_c-\mathbf{k}_p$ is the wavevector of the spin wave depending on the control and signal beams. $\mathbf{r}_j (t)$ is the position vector of the $j$-th excited atom at time $t$, related to the motion-type parameter sensing~\cite{Chen2010Heralded,Kuan2016Large,Wang2023Thermometry}. $\Delta\omega_{gs}$ is the frequency shift between the two atomic states $|g\rangle$ and $|s\rangle$ subjected to an external field. More specifically, $\Delta\omega_{gs}=\mu_{B} (m_{g} g_{g}-m_{s} g_{s} )B_z/\hbar$ is the Zeeman shift under the influence of a magnetic field $B_z$ with the Bohr magneton $\mu_{B}$, the Land\'e $g$-factors, and the magnetic sublevels $m_{g(s)}$. If, for simplicity, we let $\gamma$ be the gyromagnetic ratio instead of ${\mu}_B (m_{g} g_{g}-m_{s} g_{s} )/\hbar$, then the phase shift of the stored spin wave can be written as $\Delta\phi=\gamma B_z \Delta \tau$. In our case, $\gamma/2\pi\approx \SI{18.8}{GHz/T}$ for $^{85}\text{Rb}$ with the specific Zeeman states $|g:5^{2}S_{1/2},F=2,m_{F}=2\rangle$ and $|s:5^{2}S_{1/2},F=3,m_{F}=2\rangle$~\cite{SteckRb85DLine}. \par

For a typical MZI scheme, in order to let the signals propagating through the long and short paths combine again at a second beam splitter to form an interference fringe, an extra fiber needs to be inserted to delay the earlier signal. Taking the case with a storage time $\Delta\tau=\SI{300}{\mu s}$ as an example, a $\SI{60}{km}$ fiber delay line is required. Note that at a fiber loss rate at $\SI{795}{nm}$ of $\alpha=\SI{3}{dB/km}$, an attenuation of $\SI{180}{dB}$ is introduced. (See \hyperref[appendix]{Appendix}) Here, following the method in Ref.~\cite{Zhang2025Optical}, we employ an HD to measure the quadrature-phase amplitudes of the two signal pulses by introducing an LO, which comes from a laser injection-locked to the signal laser, beating with the signal light, as shown in Fig.~\ref{fig:setup}. The output of the $\SI{75}{MHz}$ bandwidth BHD is monitored by a fast digitizer with a \SI{100}{MHz} sampling rate, corresponding to an effective bandwidth $\SI{50}{MHz}$. By adding the recorded photocurrents from the two pulses after an electronic delay equal to the storage time, we obtain $i_+(t)$ (see \hyperref[appendix]{Appendix}). \par 

As in Fig.~\ref{fig:fringes}(a), the widths of the write and read pulses are $t_{w}=\SI{2}{\mu s}$ and $t_{r}=\SI{3}{\mu s}$, respectively, while the widths of the T and R pulses are around $\SI{1}{\mu s}$ due to the large Rabi frequency of the control light $\Omega_{c}$. The storage time $\Delta\tau$ is $\SI{5}{\mu s}$ and the total duration of each measurement window $t_{tot}$ is $\SI{10}{\mu s}$. In each cycle after MOT preparation, we apply $N=100$ measurement repetitions. We scan the atomic phase of the interferometer using a magnetic field along the $z$-direction, generated by two Helmholtz coils, with a scanning speed slower than the MOT repetition rate. To eliminate the influence of the sub-hertz frequency drift, we detune the LO with respect to the signal light by $\delta\omega/2\pi=\SI{5}{kHz}$, so that the HD becomes a heterodyne detection scheme. These recorded signals, $i_{T}(t)$, $i_{R}(t)$, and $i_{+}(t)$, for various $\Delta\phi$ induced by the magnetic field $B_{z}$, are shown in Figs.~\ref{fig:fringes}(b)-(g), in the form of pulse area integration, with (b), (d), and (f) showing a clear beat frequency $\delta\omega$ in $i_{T}(t)$, $i_{R}(t)$, and the summation $i_{+}(t)$ of T and R in (c), (e), and (g), showing destructive to constructive interference. In Fig.~\ref{fig:fringes}(h), we record the phase difference variation as a function of the applied magnetic field. Due to the random phase introduced by air flow, the phases of T (purple dots) and R (green dots) themselves show no obvious relation to the magnetic field. While the phase difference (yellow dots) between them shows a clearly linear accumulation due to the increase of the magnetic field, but it also has an ambiguity of $n\cdot2\pi$. The output power of the HD, recorded by a spectrum analyzer, is proportional to $\langle i_{+}^{2}\rangle$. As a result, we plot the power of the current addition $i_{+}$ while varying the magnetic field $B_z$, as shown as dots in Fig.~\ref{fig:fringes}(i). The recovered interference pattern is satisfactorily consistent with the fitted curve from Eq.~(\ref{eq:iave}). The difference between theory and experiment is attributed to the imperfect beating patterns. \par

If we consider the response function of the homodyne detector to be $k(t)$, the visibility is written as (see \hyperref[appendix]{Appendix}) 
\begin{equation}\label{eq:visibility}
    \begin{aligned}
    \mathcal{V}(\delta\omega)\approx\mathcal{V}_{0}\int d\omega\tilde{\mathcal{K}}(\omega)\tilde{\mathcal{A}}_{TR}(\delta\omega-\omega),
    \end{aligned}
\end{equation}
where $\tilde{\mathcal{K}}(\omega)$ and $\tilde{\mathcal{A}}_{TR}(\omega)$ are the Fourier transforms of $\mathcal{K}(\tau)=\int dt k(t)k(t+\tau)$ and $\mathcal{A}_{TR}(\tau)$, respectively. $\mathcal{A}_{TR}(\tau)=\langle a_{T}(t)a_{R}(t-\tau)\rangle$ is the pulse amplitude cross-correlation function of T and R. $\tilde{\mathcal{K}}(\omega)$ is the response spectrum of the detector, and $\tilde{\mathcal{A}}_{TR}(\omega)$ is the pulse spectrum cross-correlation. Here, we discuss two specific cases depending on the response function $k(t)$ and the pulse profile $a(t)$: (I) If the response function $k(t)$ is very slow, such that $\delta\nu_{k}\ll \delta\nu_{a}$, where $\delta\nu_{k}$ and $\delta\nu_{a}$ are the bandwidths of $\tilde{\mathcal{K}}(\omega)$ and $\tilde{\mathcal{A}}_{TR}(\omega)$, we can consider the response spectrum $\tilde{\mathcal{K}}(\omega)$ to be a delta-function; thus $\mathcal{V}(\delta\omega)\approx\mathcal{V}_{0}\tilde{\mathcal{A}}_{TR}(\delta\omega)$. (II) If the response function $k(t)$ is very fast, such that $\delta\nu_{k}\gg \delta\nu_{a}$, we can consider the pulse spectrum  $\tilde{\mathcal{A}}_{TR}(\omega)$ to be a delta-function, i.e., $\tilde{\mathcal{A}}_{TR}(\omega)\approx\mathcal{A}_{TR}(0)\delta(\omega)$; then Eq.(\ref{eq:visibility}) gives $\mathcal{V}(\delta\omega)\approx\mathcal{V}_{0}\mathcal{A}_{TR}(0)\tilde{\mathcal{K}}(\delta\omega)$. \par

\begin{figure}[htbp]
    \centering
    \includegraphics[width=1.00\linewidth]{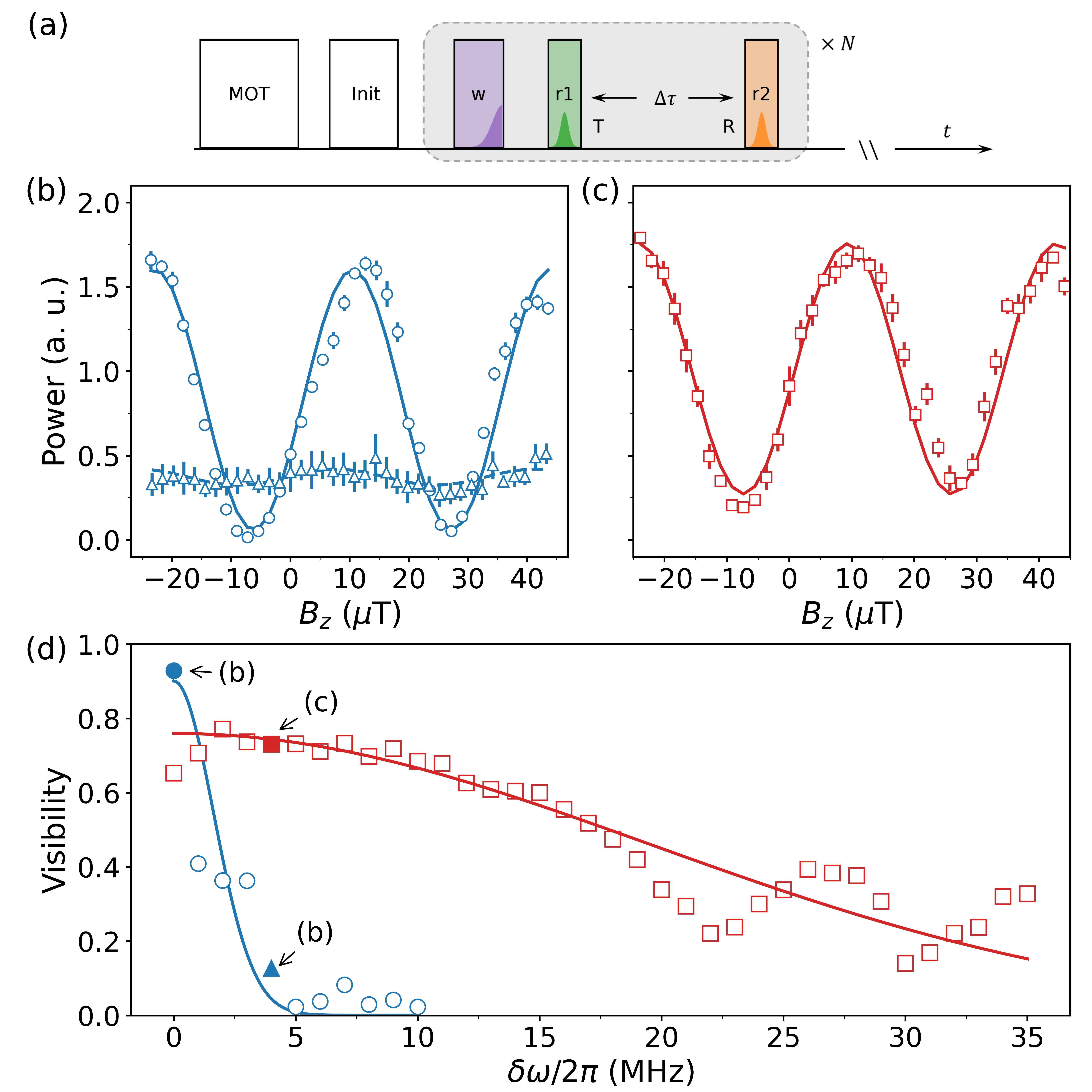}
    \caption{(a) The measurement sequence with multiple readout pulses. The first and second retrieved signals r1 and r2 are denoted as T and R in Fig.~\ref{fig:fringes}(a), and their storage time is $\Delta\tau=\SI{5}{\mu s}$. (b) The recovered interference fringe pattern obtained using a slow detector, where frequency differences $\delta\omega/2\pi=\SI{5}{kHz}$ and $\delta\omega/2\pi=\SI{4}{MHz}$ are marked as circles and triangles, respectively; the solid lines are the fitting curves. (c) The recovered interference fringe for the fast detector case, with a frequency difference $\delta\omega/2\pi=\SI{4}{MHz}$; the square points and solid line are the experimental data and the fitting curve, respectively. (d) Visibility as a function of the frequency shift of the LO, i.e., $\delta\omega$. The visibility of the curves in (b) and (c) is marked by the corresponding filled data points.}
    \label{fig:visibility}
\end{figure}

Since the data analysis method in Fig.~\ref{fig:fringes} is by pulse area integration (see \hyperref[appendix]{Appendix}), the result corresponds to the slow detector case (I), where the acceptable $\delta\omega$ between the LO and the signal field is limited by the pulse spectral width. Next, we will see that the tolerable limit extends to a large $\delta\omega$ in this protocol by using a fast detector as in case (II), which allows point-to-point addition of the photocurrent data, $i_{T}$ and $i_{R}$, leading to the result that the visibility is only limited by the response function $k(t)$ of the broadband BHD ($\SI{50}{MHz}$ in our experimental setup), according to Eq.~(\ref{eq:visibility}). In this case, the visibility is proportional to the pulse correlation function $\mathcal{A}_{TR}(0)$, which can be maximized by matching the pulse shapes of T and R. However, the T and R pulses are asymmetric, which is typical for the optical storage process~\cite{Novikova2007Optimal}. In order to obtain acceptable interference visibility, we alternatively apply a multipulse memory operation sequence with multiple readout pulses~\cite{Reim2012Multipulse}, as shown in Fig.~\ref{fig:visibility}(a). This new scheme facilitates a perfect match between the temporal profiles of the retrieved pulses; we select the first (r1) and second (r2) readout signal pulses as the data for $i_{T}$ and $i_{R}$. In this case, the width of the write pulse is kept at $t_{w}=\SI{2}{\mu s}$, and the widths of the two readout pulses are both $t_{r}=\SI{200}{ns}$. In order to maintain a satisfactory readout efficiency, the first readout pulse is applied immediately after the write pulse. The storage time between the two readout pulses $\Delta\tau$ is $\SI{5}{\mu s}$, and the total duration of each measurement is still $\SI{10}{\mu s}$. The ratio between the two readout pulses is adjusted by varying the intensity and the pulse width of the control field. \par

With this time-sequence scheme, it is now fair to compare the visibility given by Eq.~(\ref{eq:visibility}) in the two cases with a fast and a slow detector, respectively. Fig.~\ref{fig:visibility}(b) shows case (I), and the visibility of recovered interference is severely reduced when the frequency difference $\delta\omega/2\pi$ changes from $\SI{5}{kHz}$ (blue circles) to $\SI{4}{MHz}$ (blue triangles). However, for case (II), when $\delta\omega/2\pi=\SI{4}{MHz}$, visibility does not change (see \hyperref[appendix]{Appendix}) as shown in Fig.~\ref{fig:visibility}(c). To show the full trend between visibility and frequency difference more clearly, we depict them in Fig.~\ref{fig:visibility}(d) as blue and red points for case (I) and case (II), respectively. The solid curves are a Gaussian function fit to the decays of the two cases by assuming that both the response function $k(t)$ and the pulse profile $a(t)$ have a Gaussian shape. Specifically, the bandwidths of the Gaussian fits are $\delta\nu_{\mathcal{V}}^{(I)}\approx3.84\pm1.94~\SI{}{MHz}$ and $\delta\nu_{\mathcal{V}}^{(II)}\approx46.01\pm20.00~\SI{}{MHz}$, respectively. This fitting result is evidently limited by the pulse spectrum profile $\delta\nu_{a}\approx\SI{2.21}{MHz}$ and the detector response $\delta\nu_{k}\approx\SI{50}{MHz}$, as predicted by Eq.~(\ref{eq:visibility}). The labeled solid filled data points correspond to the respective cases in Figs.~\ref{fig:visibility}(b) and (c). \par

As shown in Fig.~\ref{fig:fringes} and Fig.~\ref{fig:visibility}, the interferometric phase is accumulated through the stored atomic spin wave under the influence of a magnetic field along the z-axis, $B_z$. During operation around the maximum slope $\partial \langle i_{+}^{2}(t)\rangle/\partial B_{z}$, the power variation $\text{Var}(\langle i_{+}^{2}(t)\rangle)$ determines the magnetic field uncertainty $\sigma_{B_{z}}\simeq e^{\Delta\tau/T_2^*}/(\mathcal{V}\gamma \Delta\tau\sqrt{N_{eff}})$. (See \hyperref[appendix]{Appendix}) The exponential term $e^{\Delta\tau/T_2^*}$ accounts for decoherence governed by the effective transverse coherence time $T_2^*$. The quantity $N_{eff}=\eta N_{p}$ denotes the effective photon number contributing to the signal, due to the linear mapping condition of beam-splitter-type-interaction requiring the optical excitation $N_{p}$ smaller than the number of atoms $N_{a}$, photon shot noise is dominated, and $\eta$ is the read-out efficiency that incorporates classical noise contributions. Such as the residual magnetic field induced by eddy currents from surrounding metallic components and the ambient magnetic filed that comes from the environment. 

\begin{figure}[htbp]
    \centering
    \includegraphics[width=1.00\linewidth]{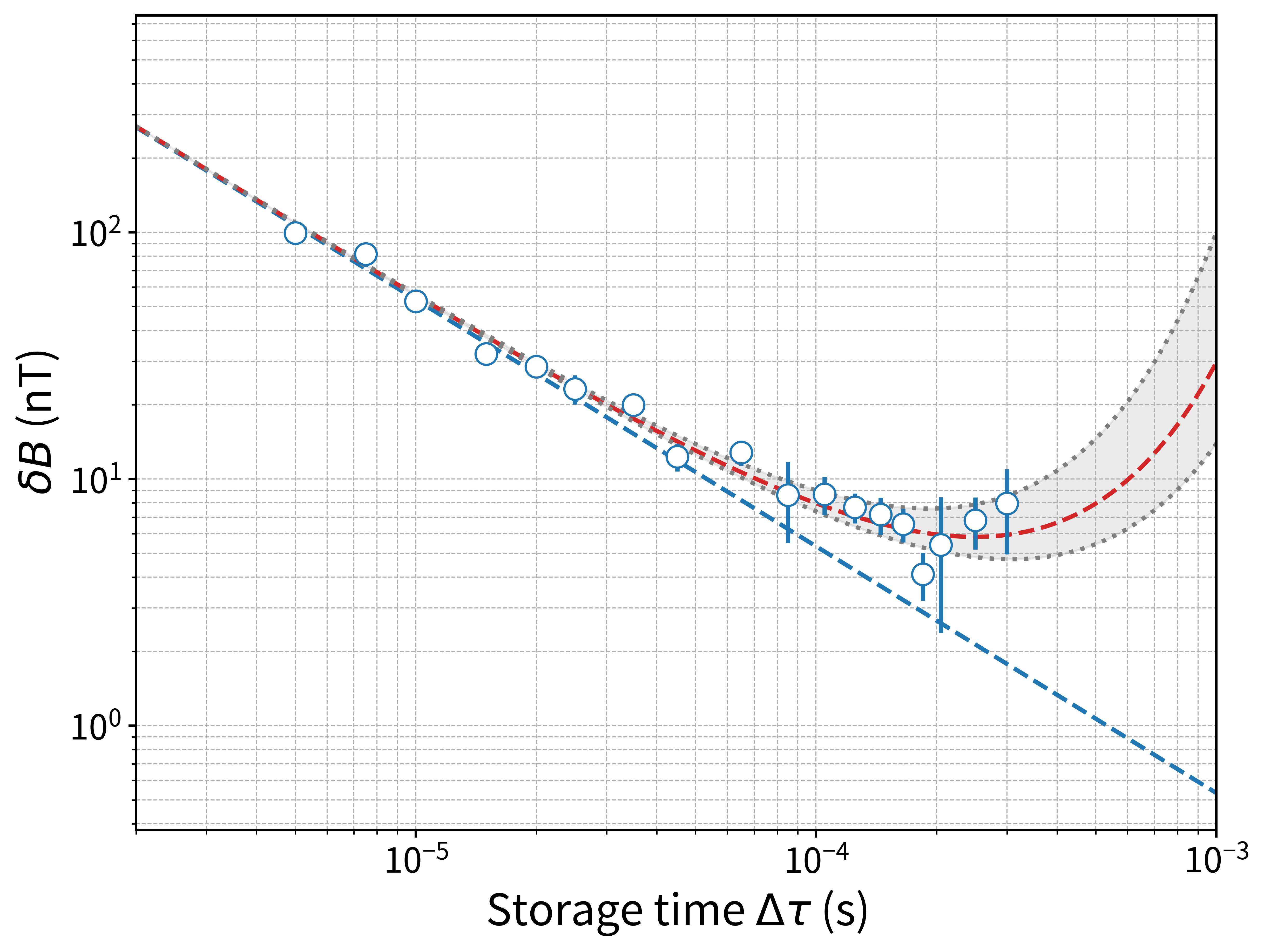}
    \caption{The magnetic field measurement precision of the interferometer versus storage time $\Delta\tau$. The circle points are the measurement data and the dashed lines (The blue dashed line $1/(\gamma\Delta\tau\sqrt{N_{eff}})$ and the red dashed line $e^{\Delta\tau/T^*_2}/(\mathcal{V}\gamma\Delta\tau\sqrt{N_{eff}})$) are the fitting curve with a $\Delta\tau^{-1}$ scaling, where $N_{eff}$ is the efficient photon number of signal field, $T_2^*$ is the coherent time of spin wave. The gray shaded region denotes $T_2^*$ in the range of $248.9 \pm 57.8~\SI{}{\mu s}$. Error bars indicate the standard deviation of all measurements. 
    }
    \label{fig:sensitivity}
\end{figure}

To observe the dependence of measurement uncertainty on storage time, we vary the storage time $\Delta\tau$ between the first and second readout pulses within the coherence lifetime. Here, we have incorporated two layers magnetic-shielding device into our experimental setup to suppress ambient magnetic field noise, and adopted a collinear propagation geometry for the signal and control beams to reduce the decoherence processes. As a results, a storage time $\SI{300}{\mu s}$ can obtained, and the measurement data are plotted in Fig.~\ref{fig:sensitivity}, during which $\delta\omega/2\pi=\SI{5}{MHz}$, the widths of two readout pulses $r_1$ and $r_2$ are $\SI{1}{\mu s}$. The cooling-and-measurement sequence was repeated over 10 cycles, with a single write-and-read measurement executed in the measurement segment of every cycle. The experiment data are fitted with $T_2^*=248.9 \pm 57.8~\SI{}{\mu s}$ and $N_{eff}=470\pm34$ (The red dashed line), by using visibility $\mathcal{V}\approx0.73$ from Fig.~\ref{fig:visibility}(d). The lowest achieved uncertainty $\delta B$ is around $\SI{6}{nT}$, for which the limitation from the ambient magnetic field is well suppressed by the ambient magnetic shield. The decoherence in our setup is mainly caused by the inhomogeneous magnetic field induced by eddy current generated after the trapping coil is switched off. According to the gyromagnetic ratio $\gamma/2\pi\approx \SI{1.88}{GHz/T}$, the measured $T_2^*$ of $248.9 \pm 57.8~\SI{}{\mu s}$ is therefore consistent with a residual inhomogeneity of about $0.3 \pm 0.1$~mG along the cigarette-shape-cloud~\cite{Wang_PhysRevResearch2024}. Decoherence due to the finite wavelength of atomic spin wave is negligible in a collinear geometry, since a atomic thermal velocity of about $\SI{10}{cm/s}$ corresponds to a dephasing time $1/(\Delta k\,v)$ with the order of $\SI{100}{ms}$.
\par

In summary, we demonstrate a new protocol for operating an atom-light hybrid interferometer for magnetic field sensing enabled by quantum storage. We experimentally show that, via HD, the interference between the transmitted and subsequently retrieved signals is recovered from data analysis of photocurrents without any time overlap between the two signals. We therefore verify that the quantum memory is an instrumental feature of our magnetometry, acting as a coherent process that leads to the accumulation of phase. Moreover, extending the storage lifetime will directly improve the sensitivity. Compared to standard atom interferometers, the atom-light interferometer compatible with quantum memory demonstrated here is predominantly limited by the photon number of the optical mode; this should be significantly smaller than the atom population, stemming from the beam-splitter-type interaction model. In a dense cold atomic ensemble utilized in faithful quantum memory, the atom number reaches $10^{9}$, and therefore a photon number of $10^{6}-10^{7}$ or higher will be guaranteed, comparable to cold atomic interferometer with effective particle number $10^{6}-10^{7}$~\cite{Ma2025Adaptive,Cohen2019Cold,cassens2025entanglement,liu2022nonlinear}. Therefore the effective atom number in this experimental platform is not a technical constrain for a single sensor node. 

Furthermore, the atom-light interferometer features both quantum memory and a beam-splitter-type interaction between atoms and light. A nonlocal interferometer composed of distributed atom-light interferometers is promising for sensing large scale physical quantities that can be converted into a phase accumulated by atoms alone (e.g., magnetic fields or gravity).
With the convenience of a photonic arm, the entangled flying photons storage in each distributed interferometer could expand to matter entanglement at a distant location \cite{Yan2017Establishing}.
The architecture of a Very-Long-Baseline Interferometer~\cite{John_D_Monnier_2003,LongBaseLine2012} is an attractive direction for building a quantum network of atomic sensors, which has been demonstrated in tabletop combining with quantum memory for imaging stars~\cite{Stas2026Entanglementassisted,Wang2026Memoryassisted}, alternatively, it could be used to measure worldwide dark matter or gravitational variations. If we assume $M$ ensembles entangled via the DLCZ protocol or linear mapping realized with quantum memories, the sensitivity of such a quantum network of probes is expected to be improved by a factor of $\sqrt{M}$ over the individual probes. The atom-light interferometer hence take the advantage of optical entanglement resource to connect distant matter nodes, to reach the ultimate Heisenberg limit~\cite{Guo2020Distributed,Xia2020Demonstration}.

\begin{acknowledgments}
This work is supported by the National Natural Science Foundation of China (NSFC) through Grants No.~92476102, No.~12404409, No.~92265109, and No.~12474262, and by the General Research Fund from Hong Kong Research Grants Council through No.~11315822 and No.~11307625. J.~F.~C. acknowledges the Guangdong Key Project under Grant No. 2022B1515020096. G.~A.~S. acknowledges support from the Hellenic Foundation for Research and Innovation (H.F.R.I.) under Project No. 11496 (``PSEUDOTOPPOS'').
X.~W. acknowledges support from the SUSTech Presidential Postdoctoral Fellowship. J.~W. acknowledges the support from City University of Hong Kong under Project No.~9610522.
\end{acknowledgments}

$^{*}$ These authors contributed equally to this work. \par
$^{\ddagger}$ jeffou@cityu.edu.hk \par
$^{\dagger}$ chenjf@sustech.edu.cn \par
$^{\S}$ Present address: Institute of Opto-Electronics, Shanxi University, Taiyuan 030006, China

\bibliography{refs} 

\begin{thebibliography}{48}%
\makeatletter
\providecommand \@ifxundefined [1]{%
 \@ifx{#1\undefined}
}%
\providecommand \@ifnum [1]{%
 \ifnum #1\expandafter \@firstoftwo
 \else \expandafter \@secondoftwo
 \fi
}%
\providecommand \@ifx [1]{%
 \ifx #1\expandafter \@firstoftwo
 \else \expandafter \@secondoftwo
 \fi
}%
\providecommand \natexlab [1]{#1}%
\providecommand \enquote  [1]{``#1''}%
\providecommand \bibnamefont  [1]{#1}%
\providecommand \bibfnamefont [1]{#1}%
\providecommand \citenamefont [1]{#1}%
\providecommand \href@noop [0]{\@secondoftwo}%
\providecommand \href [0]{\begingroup \@sanitize@url \@href}%
\providecommand \@href[1]{\@@startlink{#1}\@@href}%
\providecommand \@@href[1]{\endgroup#1\@@endlink}%
\providecommand \@sanitize@url [0]{\catcode `\\12\catcode `\$12\catcode `\&12\catcode `\#12\catcode `\^12\catcode `\_12\catcode `\%12\relax}%
\providecommand \@@startlink[1]{}%
\providecommand \@@endlink[0]{}%
\providecommand \url  [0]{\begingroup\@sanitize@url \@url }%
\providecommand \@url [1]{\endgroup\@href {#1}{\urlprefix }}%
\providecommand \urlprefix  [0]{URL }%
\providecommand \Eprint [0]{\href }%
\providecommand \doibase [0]{https://doi.org/}%
\providecommand \selectlanguage [0]{\@gobble}%
\providecommand \bibinfo  [0]{\@secondoftwo}%
\providecommand \bibfield  [0]{\@secondoftwo}%
\providecommand \translation [1]{[#1]}%
\providecommand \BibitemOpen [0]{}%
\providecommand \bibitemStop [0]{}%
\providecommand \bibitemNoStop [0]{.\EOS\space}%
\providecommand \EOS [0]{\spacefactor3000\relax}%
\providecommand \BibitemShut  [1]{\csname bibitem#1\endcsname}%
\let\auto@bib@innerbib\@empty
\bibitem [{\citenamefont {Chen}\ \emph {et~al.}(2015)\citenamefont {Chen}, \citenamefont {Qiu}, \citenamefont {Chen}, \citenamefont {Guo}, \citenamefont {Chen}, \citenamefont {Ou},\ and\ \citenamefont {Zhang}}]{ECNU_hybrid2015}%
  \BibitemOpen
  \bibfield  {author} {\bibinfo {author} {\bibfnamefont {B.}~\bibnamefont {Chen}}, \bibinfo {author} {\bibfnamefont {C.}~\bibnamefont {Qiu}}, \bibinfo {author} {\bibfnamefont {S.}~\bibnamefont {Chen}}, \bibinfo {author} {\bibfnamefont {J.}~\bibnamefont {Guo}}, \bibinfo {author} {\bibfnamefont {L.~Q.}\ \bibnamefont {Chen}}, \bibinfo {author} {\bibfnamefont {Z.~Y.}\ \bibnamefont {Ou}},\ and\ \bibinfo {author} {\bibfnamefont {W.}~\bibnamefont {Zhang}},\ }\bibfield  {title} {\bibinfo {title} {Atom-{Light} {Hybrid} {Interferometer}},\ }\href {https://doi.org/10.1103/PhysRevLett.115.043602} {\bibfield  {journal} {\bibinfo  {journal} {Phys. Rev. Lett.}\ }\textbf {\bibinfo {volume} {115}},\ \bibinfo {pages} {043602} (\bibinfo {year} {2015})}\BibitemShut {NoStop}%
\bibitem [{\citenamefont {Wen}\ \emph {et~al.}(2019)\citenamefont {Wen}, \citenamefont {Zou}, \citenamefont {Zhu}, \citenamefont {Chen}, \citenamefont {Ou}, \citenamefont {Chen},\ and\ \citenamefont {Zhang}}]{Wen2019PRL}%
  \BibitemOpen
  \bibfield  {author} {\bibinfo {author} {\bibfnamefont {R.}~\bibnamefont {Wen}}, \bibinfo {author} {\bibfnamefont {C.-L.}\ \bibnamefont {Zou}}, \bibinfo {author} {\bibfnamefont {X.}~\bibnamefont {Zhu}}, \bibinfo {author} {\bibfnamefont {P.}~\bibnamefont {Chen}}, \bibinfo {author} {\bibfnamefont {Z.~Y.}\ \bibnamefont {Ou}}, \bibinfo {author} {\bibfnamefont {J.~F.}\ \bibnamefont {Chen}},\ and\ \bibinfo {author} {\bibfnamefont {W.}~\bibnamefont {Zhang}},\ }\bibfield  {title} {\bibinfo {title} {Non-{Hermitian} {Magnon}-{Photon} {Interference} in an {Atomic} {Ensemble}},\ }\href {https://link.aps.org/doi/10.1103/PhysRevLett.122.253602} {\bibfield  {journal} {\bibinfo  {journal} {Phys. Rev. Lett.}\ }\textbf {\bibinfo {volume} {122}},\ \bibinfo {pages} {253602} (\bibinfo {year} {2019})}\BibitemShut {NoStop}%
\bibitem [{\citenamefont {Wang}\ \emph {et~al.}(2022)\citenamefont {Wang}, \citenamefont {Wang}, \citenamefont {Ren}, \citenamefont {Wen}, \citenamefont {Zou}, \citenamefont {Siviloglou},\ and\ \citenamefont {Chen}}]{Wang2022Quantum}%
  \BibitemOpen
  \bibfield  {author} {\bibinfo {author} {\bibfnamefont {X.}~\bibnamefont {Wang}}, \bibinfo {author} {\bibfnamefont {J.}~\bibnamefont {Wang}}, \bibinfo {author} {\bibfnamefont {Z.}~\bibnamefont {Ren}}, \bibinfo {author} {\bibfnamefont {R.}~\bibnamefont {Wen}}, \bibinfo {author} {\bibfnamefont {C.-L.}\ \bibnamefont {Zou}}, \bibinfo {author} {\bibfnamefont {G.~A.}\ \bibnamefont {Siviloglou}},\ and\ \bibinfo {author} {\bibfnamefont {J.~F.}\ \bibnamefont {Chen}},\ }\bibfield  {title} {\bibinfo {title} {Quantum {{Interference}} between {{Photons}} and {{Single Quanta}} of {{Stored Atomic Coherence}}},\ }\href {https://doi.org/10.1103/PhysRevLett.128.083605} {\bibfield  {journal} {\bibinfo  {journal} {Phys. Rev. Lett.}\ }\textbf {\bibinfo {volume} {128}},\ \bibinfo {pages} {083605} (\bibinfo {year} {2022})}\BibitemShut {NoStop}%
\bibitem [{\citenamefont {Su}\ \emph {et~al.}(2022)\citenamefont {Su}, \citenamefont {Zhong}, \citenamefont {Zhang}, \citenamefont {Li}, \citenamefont {Zou}, \citenamefont {Wang}, \citenamefont {Yan},\ and\ \citenamefont {Zhu}}]{Su2022Quantum}%
  \BibitemOpen
  \bibfield  {author} {\bibinfo {author} {\bibfnamefont {K.}~\bibnamefont {Su}}, \bibinfo {author} {\bibfnamefont {Y.}~\bibnamefont {Zhong}}, \bibinfo {author} {\bibfnamefont {S.}~\bibnamefont {Zhang}}, \bibinfo {author} {\bibfnamefont {J.}~\bibnamefont {Li}}, \bibinfo {author} {\bibfnamefont {C.-L.}\ \bibnamefont {Zou}}, \bibinfo {author} {\bibfnamefont {Y.}~\bibnamefont {Wang}}, \bibinfo {author} {\bibfnamefont {H.}~\bibnamefont {Yan}},\ and\ \bibinfo {author} {\bibfnamefont {S.-L.}\ \bibnamefont {Zhu}},\ }\bibfield  {title} {\bibinfo {title} {Quantum interference between nonidentical single particles},\ }\href {https://doi.org/10.1103/PhysRevLett.129.093604} {\bibfield  {journal} {\bibinfo  {journal} {Phys. Rev. Lett.}\ }\textbf {\bibinfo {volume} {129}},\ \bibinfo {pages} {093604} (\bibinfo {year} {2022})}\BibitemShut {NoStop}%
\bibitem [{\citenamefont {Chen}\ \emph {et~al.}(2010)\citenamefont {Chen}, \citenamefont {Bao}, \citenamefont {Yuan}, \citenamefont {Chen}, \citenamefont {Zhao},\ and\ \citenamefont {Pan}}]{Chen2010Heralded}%
  \BibitemOpen
  \bibfield  {author} {\bibinfo {author} {\bibfnamefont {Y.-A.}\ \bibnamefont {Chen}}, \bibinfo {author} {\bibfnamefont {X.-H.}\ \bibnamefont {Bao}}, \bibinfo {author} {\bibfnamefont {Z.-S.}\ \bibnamefont {Yuan}}, \bibinfo {author} {\bibfnamefont {S.}~\bibnamefont {Chen}}, \bibinfo {author} {\bibfnamefont {B.}~\bibnamefont {Zhao}},\ and\ \bibinfo {author} {\bibfnamefont {J.-W.}\ \bibnamefont {Pan}},\ }\bibfield  {title} {\bibinfo {title} {{Heralded Generation} of an {Atomic NOON State}},\ }\href {https://doi.org/10.1103/PhysRevLett.104.043601} {\bibfield  {journal} {\bibinfo  {journal} {Phys. Rev. Lett.}\ }\textbf {\bibinfo {volume} {104}},\ \bibinfo {pages} {043601} (\bibinfo {year} {2010})}\BibitemShut {NoStop}%
\bibitem [{\citenamefont {Kuan}\ \emph {et~al.}(2016)\citenamefont {Kuan}, \citenamefont {Huang}, \citenamefont {Chan}, \citenamefont {Kosen},\ and\ \citenamefont {Lan}}]{Kuan2016Large}%
  \BibitemOpen
  \bibfield  {author} {\bibinfo {author} {\bibfnamefont {P.-C.}\ \bibnamefont {Kuan}}, \bibinfo {author} {\bibfnamefont {C.}~\bibnamefont {Huang}}, \bibinfo {author} {\bibfnamefont {W.~S.}\ \bibnamefont {Chan}}, \bibinfo {author} {\bibfnamefont {S.}~\bibnamefont {Kosen}},\ and\ \bibinfo {author} {\bibfnamefont {S.-Y.}\ \bibnamefont {Lan}},\ }\bibfield  {title} {\bibinfo {title} {Large {{Fizeau}}'s light-dragging effect in a moving electromagnetically induced transparent medium},\ }\href {https://doi.org/10.1038/ncomms13030} {\bibfield  {journal} {\bibinfo  {journal} {Nat. Commun.}\ }\textbf {\bibinfo {volume} {7}},\ \bibinfo {pages} {13030} (\bibinfo {year} {2016})}\BibitemShut {NoStop}%
\bibitem [{\citenamefont {Gomez}\ \emph {et~al.}(2020)\citenamefont {Gomez}, \citenamefont {Martin}, \citenamefont {Mazzinghi}, \citenamefont {Benedicto~Orenes}, \citenamefont {Palacios},\ and\ \citenamefont {Mitchell}}]{Gomez2020Bose}%
  \BibitemOpen
  \bibfield  {author} {\bibinfo {author} {\bibfnamefont {P.}~\bibnamefont {Gomez}}, \bibinfo {author} {\bibfnamefont {F.}~\bibnamefont {Martin}}, \bibinfo {author} {\bibfnamefont {C.}~\bibnamefont {Mazzinghi}}, \bibinfo {author} {\bibfnamefont {D.}~\bibnamefont {Benedicto~Orenes}}, \bibinfo {author} {\bibfnamefont {S.}~\bibnamefont {Palacios}},\ and\ \bibinfo {author} {\bibfnamefont {M.~W.}\ \bibnamefont {Mitchell}},\ }\bibfield  {title} {\bibinfo {title} {{Bose-Einstein Condensate Comagnetometer}},\ }\href {https://doi.org/10.1103/PhysRevLett.124.170401} {\bibfield  {journal} {\bibinfo  {journal} {Phys. Rev. Lett.}\ }\textbf {\bibinfo {volume} {124}},\ \bibinfo {pages} {170401} (\bibinfo {year} {2020})}\BibitemShut {NoStop}%
\bibitem [{\citenamefont {Yang}\ \emph {et~al.}(2025)\citenamefont {Yang}, \citenamefont {Luo}, \citenamefont {Zhang}, \citenamefont {Wang}, \citenamefont {Zou}, \citenamefont {Xia},\ and\ \citenamefont {Lu}}]{ZTLuNP_Cat_2025}%
  \BibitemOpen
  \bibfield  {author} {\bibinfo {author} {\bibfnamefont {Y.~A.}\ \bibnamefont {Yang}}, \bibinfo {author} {\bibfnamefont {W.-T.}\ \bibnamefont {Luo}}, \bibinfo {author} {\bibfnamefont {J.-L.}\ \bibnamefont {Zhang}}, \bibinfo {author} {\bibfnamefont {S.-Z.}\ \bibnamefont {Wang}}, \bibinfo {author} {\bibfnamefont {C.-L.}\ \bibnamefont {Zou}}, \bibinfo {author} {\bibfnamefont {T.}~\bibnamefont {Xia}},\ and\ \bibinfo {author} {\bibfnamefont {Z.-T.}\ \bibnamefont {Lu}},\ }\bibfield  {title} {\bibinfo {title} {Minute-scale {S}chr\"odinger-cat state of spin-5/2 atoms},\ }\href {https://doi.org/10.1038/s41566-024-01555-3} {\bibfield  {journal} {\bibinfo  {journal} {Nat. Photonics}\ }\textbf {\bibinfo {volume} {19}},\ \bibinfo {pages} {89} (\bibinfo {year} {2025})}\BibitemShut {NoStop}%
\bibitem [{\citenamefont {Ma}\ \emph {et~al.}(2025)\citenamefont {Ma}, \citenamefont {Han}, \citenamefont {Tan}, \citenamefont {He}, \citenamefont {Shi}, \citenamefont {Kang}, \citenamefont {Wu}, \citenamefont {Huang}, \citenamefont {Lu},\ and\ \citenamefont {Lee}}]{Ma2025Adaptive}%
  \BibitemOpen
  \bibfield  {author} {\bibinfo {author} {\bibfnamefont {Z.}~\bibnamefont {Ma}}, \bibinfo {author} {\bibfnamefont {C.}~\bibnamefont {Han}}, \bibinfo {author} {\bibfnamefont {Z.}~\bibnamefont {Tan}}, \bibinfo {author} {\bibfnamefont {H.}~\bibnamefont {He}}, \bibinfo {author} {\bibfnamefont {S.}~\bibnamefont {Shi}}, \bibinfo {author} {\bibfnamefont {X.}~\bibnamefont {Kang}}, \bibinfo {author} {\bibfnamefont {J.}~\bibnamefont {Wu}}, \bibinfo {author} {\bibfnamefont {J.}~\bibnamefont {Huang}}, \bibinfo {author} {\bibfnamefont {B.}~\bibnamefont {Lu}},\ and\ \bibinfo {author} {\bibfnamefont {C.}~\bibnamefont {Lee}},\ }\bibfield  {title} {\bibinfo {title} {Adaptive cold-atom magnetometry mitigating the trade-off between sensitivity and dynamic range},\ }\href {https://doi.org/10.1126/sciadv.adt3938} {\bibfield  {journal} {\bibinfo  {journal} {Sci. Adv.}\ }\textbf {\bibinfo {volume} {11}},\ \bibinfo {pages} {eadt3938} (\bibinfo {year} {2025})}\BibitemShut {NoStop}%
\bibitem [{\citenamefont {Zhao}\ \emph {et~al.}(2009{\natexlab{a}})\citenamefont {Zhao}, \citenamefont {Chen}, \citenamefont {Bao}, \citenamefont {Strassel}, \citenamefont {Chuu}, \citenamefont {Jin}, \citenamefont {Schmiedmayer}, \citenamefont {Yuan}, \citenamefont {Chen},\ and\ \citenamefont {Pan}}]{Zhao2009Millisecond}%
  \BibitemOpen
  \bibfield  {author} {\bibinfo {author} {\bibfnamefont {B.}~\bibnamefont {Zhao}}, \bibinfo {author} {\bibfnamefont {Y.-A.}\ \bibnamefont {Chen}}, \bibinfo {author} {\bibfnamefont {X.-H.}\ \bibnamefont {Bao}}, \bibinfo {author} {\bibfnamefont {T.}~\bibnamefont {Strassel}}, \bibinfo {author} {\bibfnamefont {C.-S.}\ \bibnamefont {Chuu}}, \bibinfo {author} {\bibfnamefont {X.-M.}\ \bibnamefont {Jin}}, \bibinfo {author} {\bibfnamefont {J.}~\bibnamefont {Schmiedmayer}}, \bibinfo {author} {\bibfnamefont {Z.-S.}\ \bibnamefont {Yuan}}, \bibinfo {author} {\bibfnamefont {S.}~\bibnamefont {Chen}},\ and\ \bibinfo {author} {\bibfnamefont {J.-W.}\ \bibnamefont {Pan}},\ }\bibfield  {title} {\bibinfo {title} {A millisecond quantum memory for scalable quantum networks},\ }\href {https://doi.org/10.1038/nphys1153} {\bibfield  {journal} {\bibinfo  {journal} {Nat. Phys.}\ }\textbf {\bibinfo {volume} {5}},\ \bibinfo {pages} {95} (\bibinfo {year} {2009}{\natexlab{a}})}\BibitemShut {NoStop}%
\bibitem [{\citenamefont {Zhao}\ \emph {et~al.}(2009{\natexlab{b}})\citenamefont {Zhao}, \citenamefont {Dudin}, \citenamefont {Jenkins}, \citenamefont {Campbell}, \citenamefont {Matsukevich}, \citenamefont {Kennedy},\ and\ \citenamefont {Kuzmich}}]{Zhao2009Longlived}%
  \BibitemOpen
  \bibfield  {author} {\bibinfo {author} {\bibfnamefont {R.}~\bibnamefont {Zhao}}, \bibinfo {author} {\bibfnamefont {Y.~O.}\ \bibnamefont {Dudin}}, \bibinfo {author} {\bibfnamefont {S.~D.}\ \bibnamefont {Jenkins}}, \bibinfo {author} {\bibfnamefont {C.~J.}\ \bibnamefont {Campbell}}, \bibinfo {author} {\bibfnamefont {D.~N.}\ \bibnamefont {Matsukevich}}, \bibinfo {author} {\bibfnamefont {T.~A.~B.}\ \bibnamefont {Kennedy}},\ and\ \bibinfo {author} {\bibfnamefont {A.}~\bibnamefont {Kuzmich}},\ }\bibfield  {title} {\bibinfo {title} {Long-lived quantum memory},\ }\href {https://doi.org/10.1038/nphys1152} {\bibfield  {journal} {\bibinfo  {journal} {Nat. Phys.}\ }\textbf {\bibinfo {volume} {5}},\ \bibinfo {pages} {100} (\bibinfo {year} {2009}{\natexlab{b}})}\BibitemShut {NoStop}%
\bibitem [{\citenamefont {Dudin}\ \emph {et~al.}(2013)\citenamefont {Dudin}, \citenamefont {Li},\ and\ \citenamefont {Kuzmich}}]{Dudin2013Light}%
  \BibitemOpen
  \bibfield  {author} {\bibinfo {author} {\bibfnamefont {Y.~O.}\ \bibnamefont {Dudin}}, \bibinfo {author} {\bibfnamefont {L.}~\bibnamefont {Li}},\ and\ \bibinfo {author} {\bibfnamefont {A.}~\bibnamefont {Kuzmich}},\ }\bibfield  {title} {\bibinfo {title} {Light storage on the time scale of a minute},\ }\href {https://doi.org/10.1103/PhysRevA.87.031801} {\bibfield  {journal} {\bibinfo  {journal} {Phys. Rev. A}\ }\textbf {\bibinfo {volume} {87}},\ \bibinfo {pages} {031801} (\bibinfo {year} {2013})}\BibitemShut {NoStop}%
\bibitem [{\citenamefont {Yang}\ \emph {et~al.}(2016)\citenamefont {Yang}, \citenamefont {Wang}, \citenamefont {Bao},\ and\ \citenamefont {Pan}}]{Yang2016Efficient}%
  \BibitemOpen
  \bibfield  {author} {\bibinfo {author} {\bibfnamefont {S.-J.}\ \bibnamefont {Yang}}, \bibinfo {author} {\bibfnamefont {X.-J.}\ \bibnamefont {Wang}}, \bibinfo {author} {\bibfnamefont {X.-H.}\ \bibnamefont {Bao}},\ and\ \bibinfo {author} {\bibfnamefont {J.-W.}\ \bibnamefont {Pan}},\ }\bibfield  {title} {\bibinfo {title} {An efficient quantum light--matter interface with sub-second lifetime},\ }\href {https://doi.org/10.1038/nphoton.2016.51} {\bibfield  {journal} {\bibinfo  {journal} {Nat. Photonics}\ }\textbf {\bibinfo {volume} {10}},\ \bibinfo {pages} {381} (\bibinfo {year} {2016})}\BibitemShut {NoStop}%
\bibitem [{\citenamefont {Cho}\ \emph {et~al.}(2016)\citenamefont {Cho}, \citenamefont {Campbell}, \citenamefont {Everett}, \citenamefont {Bernu}, \citenamefont {Higginbottom}, \citenamefont {Cao}, \citenamefont {Geng}, \citenamefont {Robins}, \citenamefont {Lam},\ and\ \citenamefont {Buchler}}]{Cho2016Highly}%
  \BibitemOpen
  \bibfield  {author} {\bibinfo {author} {\bibfnamefont {Y.-W.}\ \bibnamefont {Cho}}, \bibinfo {author} {\bibfnamefont {G.~T.}\ \bibnamefont {Campbell}}, \bibinfo {author} {\bibfnamefont {J.~L.}\ \bibnamefont {Everett}}, \bibinfo {author} {\bibfnamefont {J.}~\bibnamefont {Bernu}}, \bibinfo {author} {\bibfnamefont {D.~B.}\ \bibnamefont {Higginbottom}}, \bibinfo {author} {\bibfnamefont {M.~T.}\ \bibnamefont {Cao}}, \bibinfo {author} {\bibfnamefont {J.}~\bibnamefont {Geng}}, \bibinfo {author} {\bibfnamefont {N.~P.}\ \bibnamefont {Robins}}, \bibinfo {author} {\bibfnamefont {P.~K.}\ \bibnamefont {Lam}},\ and\ \bibinfo {author} {\bibfnamefont {B.~C.}\ \bibnamefont {Buchler}},\ }\bibfield  {title} {\bibinfo {title} {Highly efficient optical quantum memory with long coherence time in cold atoms},\ }\href {https://doi.org/10.1364/OPTICA.3.000100} {\bibfield  {journal} {\bibinfo  {journal} {Optica}\ }\textbf {\bibinfo {volume} {3}},\ \bibinfo {pages} {100} (\bibinfo {year} {2016})}\BibitemShut {NoStop}%
\bibitem [{\citenamefont {Hsiao}\ \emph {et~al.}(2018)\citenamefont {Hsiao}, \citenamefont {Tsai}, \citenamefont {Chen}, \citenamefont {Lin}, \citenamefont {Hung}, \citenamefont {Lee}, \citenamefont {Chen}, \citenamefont {Chen}, \citenamefont {Yu},\ and\ \citenamefont {Chen}}]{Hsiao2018Highly}%
  \BibitemOpen
  \bibfield  {author} {\bibinfo {author} {\bibfnamefont {Y.-F.}\ \bibnamefont {Hsiao}}, \bibinfo {author} {\bibfnamefont {P.-J.}\ \bibnamefont {Tsai}}, \bibinfo {author} {\bibfnamefont {H.-S.}\ \bibnamefont {Chen}}, \bibinfo {author} {\bibfnamefont {S.-X.}\ \bibnamefont {Lin}}, \bibinfo {author} {\bibfnamefont {C.-C.}\ \bibnamefont {Hung}}, \bibinfo {author} {\bibfnamefont {C.-H.}\ \bibnamefont {Lee}}, \bibinfo {author} {\bibfnamefont {Y.-H.}\ \bibnamefont {Chen}}, \bibinfo {author} {\bibfnamefont {Y.-F.}\ \bibnamefont {Chen}}, \bibinfo {author} {\bibfnamefont {I.~A.}\ \bibnamefont {Yu}},\ and\ \bibinfo {author} {\bibfnamefont {Y.-C.}\ \bibnamefont {Chen}},\ }\bibfield  {title} {\bibinfo {title} {Highly {{Efficient Coherent Optical Memory Based}} on {{Electromagnetically Induced Transparency}}},\ }\href {https://doi.org/10.1103/PhysRevLett.120.183602} {\bibfield  {journal} {\bibinfo  {journal} {Phys. Rev. Lett.}\ }\textbf {\bibinfo {volume} {120}},\ \bibinfo {pages} {183602} (\bibinfo {year}
  {2018})}\BibitemShut {NoStop}%
\bibitem [{\citenamefont {Erdmann}\ \emph {et~al.}(2000)\citenamefont {Erdmann}, \citenamefont {Branning}, \citenamefont {Grice},\ and\ \citenamefont {Walmsley}}]{Erdmann2000Restoring}%
  \BibitemOpen
  \bibfield  {author} {\bibinfo {author} {\bibfnamefont {R.}~\bibnamefont {Erdmann}}, \bibinfo {author} {\bibfnamefont {D.}~\bibnamefont {Branning}}, \bibinfo {author} {\bibfnamefont {W.}~\bibnamefont {Grice}},\ and\ \bibinfo {author} {\bibfnamefont {I.~A.}\ \bibnamefont {Walmsley}},\ }\bibfield  {title} {\bibinfo {title} {Restoring dispersion cancellation for entangled photons produced by ultrashort pulses},\ }\href {https://doi.org/10.1103/PhysRevA.62.053810} {\bibfield  {journal} {\bibinfo  {journal} {Phys. Rev. A}\ }\textbf {\bibinfo {volume} {62}},\ \bibinfo {pages} {053810} (\bibinfo {year} {2000})}\BibitemShut {NoStop}%
\bibitem [{\citenamefont {Yuen}\ and\ \citenamefont {Shapiro}(1978)}]{Yuen1978Optical}%
  \BibitemOpen
  \bibfield  {author} {\bibinfo {author} {\bibfnamefont {H.}~\bibnamefont {Yuen}}\ and\ \bibinfo {author} {\bibfnamefont {J.}~\bibnamefont {Shapiro}},\ }\bibfield  {title} {\bibinfo {title} {Optical communication with two-photon coherent states--{{Part I}}: {{Quantum-state}} propagation and quantum-noise},\ }\href {https://doi.org/10.1109/TIT.1978.1055958} {\bibfield  {journal} {\bibinfo  {journal} {IEEE Trans. Inf. Theory}\ }\textbf {\bibinfo {volume} {24}},\ \bibinfo {pages} {657} (\bibinfo {year} {1978})}\BibitemShut {NoStop}%
\bibitem [{\citenamefont {Zhang}\ \emph {et~al.}(2025)\citenamefont {Zhang}, \citenamefont {Tang}, \citenamefont {Guo}, \citenamefont {Cui}, \citenamefont {Li},\ and\ \citenamefont {Ou}}]{Zhang2025Optical}%
  \BibitemOpen
  \bibfield  {author} {\bibinfo {author} {\bibfnamefont {Y.}~\bibnamefont {Zhang}}, \bibinfo {author} {\bibfnamefont {X.}~\bibnamefont {Tang}}, \bibinfo {author} {\bibfnamefont {X.}~\bibnamefont {Guo}}, \bibinfo {author} {\bibfnamefont {L.}~\bibnamefont {Cui}}, \bibinfo {author} {\bibfnamefont {X.}~\bibnamefont {Li}},\ and\ \bibinfo {author} {\bibfnamefont {Z.~Y.}\ \bibnamefont {Ou}},\ }\bibfield  {title} {\bibinfo {title} {Optical interference by amplitude measurement},\ }\href {https://doi.org/10.1103/PhysRevResearch.7.013255} {\bibfield  {journal} {\bibinfo  {journal} {Phys. Rev. Research}\ }\textbf {\bibinfo {volume} {7}},\ \bibinfo {pages} {013255} (\bibinfo {year} {2025})}\BibitemShut {NoStop}%
\bibitem [{\citenamefont {Tang}\ \emph {et~al.}(2025)\citenamefont {Tang}, \citenamefont {Zhang}, \citenamefont {Guo}, \citenamefont {Cui}, \citenamefont {Li},\ and\ \citenamefont {Ou}}]{Tang2025Phasedependent}%
  \BibitemOpen
  \bibfield  {author} {\bibinfo {author} {\bibfnamefont {X.}~\bibnamefont {Tang}}, \bibinfo {author} {\bibfnamefont {Y.}~\bibnamefont {Zhang}}, \bibinfo {author} {\bibfnamefont {X.}~\bibnamefont {Guo}}, \bibinfo {author} {\bibfnamefont {L.}~\bibnamefont {Cui}}, \bibinfo {author} {\bibfnamefont {X.}~\bibnamefont {Li}},\ and\ \bibinfo {author} {\bibfnamefont {Z.~Y.}\ \bibnamefont {Ou}},\ }\bibfield  {title} {\bibinfo {title} {Phase-dependent {{Hanbury-Brown}} and {{Twiss}} effect for the complete measurement of the complex coherence function},\ }\href {https://doi.org/10.1038/s41377-024-01684-y} {\bibfield  {journal} {\bibinfo  {journal} {Light Sci. Appl.}\ }\textbf {\bibinfo {volume} {14}},\ \bibinfo {pages} {46} (\bibinfo {year} {2025})}\BibitemShut {NoStop}%
\bibitem [{\citenamefont {Xiao}\ \emph {et~al.}(1995)\citenamefont {Xiao}, \citenamefont {Li}, \citenamefont {Jin},\ and\ \citenamefont {{Gea-Banacloche}}}]{Xiao1995Measurement}%
  \BibitemOpen
  \bibfield  {author} {\bibinfo {author} {\bibfnamefont {M.}~\bibnamefont {Xiao}}, \bibinfo {author} {\bibfnamefont {Y.-Q.}\ \bibnamefont {Li}}, \bibinfo {author} {\bibfnamefont {S.-Z.}\ \bibnamefont {Jin}},\ and\ \bibinfo {author} {\bibfnamefont {J.}~\bibnamefont {{Gea-Banacloche}}},\ }\bibfield  {title} {\bibinfo {title} {Measurement of {{Dispersive Properties}} of {{Electromagnetically Induced Transparency}} in {{Rubidium Atoms}}},\ }\href {https://doi.org/10.1103/PhysRevLett.74.666} {\bibfield  {journal} {\bibinfo  {journal} {Phys. Rev. Lett.}\ }\textbf {\bibinfo {volume} {74}},\ \bibinfo {pages} {666} (\bibinfo {year} {1995})}\BibitemShut {NoStop}%
\bibitem [{\citenamefont {Zibrov}\ \emph {et~al.}(1996)\citenamefont {Zibrov}, \citenamefont {Lukin}, \citenamefont {Hollberg}, \citenamefont {Nikonov}, \citenamefont {Scully}, \citenamefont {Robinson},\ and\ \citenamefont {Velichansky}}]{Zibrov1996Experimental}%
  \BibitemOpen
  \bibfield  {author} {\bibinfo {author} {\bibfnamefont {A.~S.}\ \bibnamefont {Zibrov}}, \bibinfo {author} {\bibfnamefont {M.~D.}\ \bibnamefont {Lukin}}, \bibinfo {author} {\bibfnamefont {L.}~\bibnamefont {Hollberg}}, \bibinfo {author} {\bibfnamefont {D.~E.}\ \bibnamefont {Nikonov}}, \bibinfo {author} {\bibfnamefont {M.~O.}\ \bibnamefont {Scully}}, \bibinfo {author} {\bibfnamefont {H.~G.}\ \bibnamefont {Robinson}},\ and\ \bibinfo {author} {\bibfnamefont {V.~L.}\ \bibnamefont {Velichansky}},\ }\bibfield  {title} {\bibinfo {title} {Experimental {{Demonstration}} of {{Enhanced Index}} of {{Refraction}} via {{Quantum Coherence}} in {{Rb}}},\ }\href {https://doi.org/10.1103/PhysRevLett.76.3935} {\bibfield  {journal} {\bibinfo  {journal} {Phys. Rev. Lett.}\ }\textbf {\bibinfo {volume} {76}},\ \bibinfo {pages} {3935} (\bibinfo {year} {1996})}\BibitemShut {NoStop}%
\bibitem [{\citenamefont {Aljunid}\ \emph {et~al.}(2009)\citenamefont {Aljunid}, \citenamefont {Tey}, \citenamefont {Chng}, \citenamefont {Liew}, \citenamefont {Maslennikov}, \citenamefont {Scarani},\ and\ \citenamefont {Kurtsiefer}}]{Aljunid2009Phase}%
  \BibitemOpen
  \bibfield  {author} {\bibinfo {author} {\bibfnamefont {S.~A.}\ \bibnamefont {Aljunid}}, \bibinfo {author} {\bibfnamefont {M.~K.}\ \bibnamefont {Tey}}, \bibinfo {author} {\bibfnamefont {B.}~\bibnamefont {Chng}}, \bibinfo {author} {\bibfnamefont {T.}~\bibnamefont {Liew}}, \bibinfo {author} {\bibfnamefont {G.}~\bibnamefont {Maslennikov}}, \bibinfo {author} {\bibfnamefont {V.}~\bibnamefont {Scarani}},\ and\ \bibinfo {author} {\bibfnamefont {C.}~\bibnamefont {Kurtsiefer}},\ }\bibfield  {title} {\bibinfo {title} {Phase {{Shift}} of a {{Weak Coherent Beam Induced}} by a {{Single Atom}}},\ }\href {https://doi.org/10.1103/PhysRevLett.103.153601} {\bibfield  {journal} {\bibinfo  {journal} {Phys. Rev. Lett.}\ }\textbf {\bibinfo {volume} {103}},\ \bibinfo {pages} {153601} (\bibinfo {year} {2009})}\BibitemShut {NoStop}%
\bibitem [{\citenamefont {Appel}\ \emph {et~al.}(2008)\citenamefont {Appel}, \citenamefont {Figueroa}, \citenamefont {Korystov}, \citenamefont {Lobino},\ and\ \citenamefont {Lvovsky}}]{Appel2008Quantum}%
  \BibitemOpen
  \bibfield  {author} {\bibinfo {author} {\bibfnamefont {J.}~\bibnamefont {Appel}}, \bibinfo {author} {\bibfnamefont {E.}~\bibnamefont {Figueroa}}, \bibinfo {author} {\bibfnamefont {D.}~\bibnamefont {Korystov}}, \bibinfo {author} {\bibfnamefont {M.}~\bibnamefont {Lobino}},\ and\ \bibinfo {author} {\bibfnamefont {A.~I.}\ \bibnamefont {Lvovsky}},\ }\bibfield  {title} {\bibinfo {title} {Quantum memory for squeezed light},\ }\href {https://doi.org/10.1103/PhysRevLett.100.093602} {\bibfield  {journal} {\bibinfo  {journal} {Phys. Rev. Lett.}\ }\textbf {\bibinfo {volume} {100}},\ \bibinfo {pages} {093602} (\bibinfo {year} {2008})}\BibitemShut {NoStop}%
\bibitem [{\citenamefont {Giovannetti}\ \emph {et~al.}(2004)\citenamefont {Giovannetti}, \citenamefont {Lloyd},\ and\ \citenamefont {Maccone}}]{doi:10.1126/science.1104149}%
  \BibitemOpen
  \bibfield  {author} {\bibinfo {author} {\bibfnamefont {V.}~\bibnamefont {Giovannetti}}, \bibinfo {author} {\bibfnamefont {S.}~\bibnamefont {Lloyd}},\ and\ \bibinfo {author} {\bibfnamefont {L.}~\bibnamefont {Maccone}},\ }\bibfield  {title} {\bibinfo {title} {Quantum-enhanced measurements: Beating the standard quantum limit},\ }\href {https://doi.org/10.1126/science.1104149} {\bibfield  {journal} {\bibinfo  {journal} {Science}\ }\textbf {\bibinfo {volume} {306}},\ \bibinfo {pages} {1330} (\bibinfo {year} {2004})}\BibitemShut {NoStop}%
\bibitem [{\citenamefont {Guo}\ \emph {et~al.}(2020{\natexlab{a}})\citenamefont {Guo}, \citenamefont {Breum}, \citenamefont {Borregaard}, \citenamefont {Izumi}, \citenamefont {Larsen}, \citenamefont {Gehring}, \citenamefont {Christandl}, \citenamefont {Neergaard-Nielsen},\ and\ \citenamefont {L.}}]{XueshiNP2020}%
  \BibitemOpen
  \bibfield  {author} {\bibinfo {author} {\bibfnamefont {X.}~\bibnamefont {Guo}}, \bibinfo {author} {\bibfnamefont {C.~R.}\ \bibnamefont {Breum}}, \bibinfo {author} {\bibfnamefont {J.}~\bibnamefont {Borregaard}}, \bibinfo {author} {\bibfnamefont {S.}~\bibnamefont {Izumi}}, \bibinfo {author} {\bibfnamefont {M.~V.}\ \bibnamefont {Larsen}}, \bibinfo {author} {\bibfnamefont {T.}~\bibnamefont {Gehring}}, \bibinfo {author} {\bibfnamefont {M.}~\bibnamefont {Christandl}}, \bibinfo {author} {\bibfnamefont {J.~S.}\ \bibnamefont {Neergaard-Nielsen}},\ and\ \bibinfo {author} {\bibfnamefont {A.~U.}\ \bibnamefont {L.}},\ }\bibfield  {title} {\bibinfo {title} {Distributed quantum sensing in a continuous-variable entangled network},\ }\href {https://doi.org/10.1038/s41567-019-0743-x} {\bibfield  {journal} {\bibinfo  {journal} {Nat. Phys.}\ }\textbf {\bibinfo {volume} {16}},\ \bibinfo {pages} {281} (\bibinfo {year} {2020}{\natexlab{a}})}\BibitemShut {NoStop}%
\bibitem [{\citenamefont {Liu}\ \emph {et~al.}(2021)\citenamefont {Liu}, \citenamefont {Zhang}, \citenamefont {Li}, \citenamefont {Zhang}, \citenamefont {Yin}, \citenamefont {Fei}, \citenamefont {Li}, \citenamefont {Liu}, \citenamefont {Xu}, \citenamefont {Chen},\ and\ \citenamefont {Pan}}]{yuao2021NPdistributed}%
  \BibitemOpen
  \bibfield  {author} {\bibinfo {author} {\bibfnamefont {L.-Z.}\ \bibnamefont {Liu}}, \bibinfo {author} {\bibfnamefont {Y.-Z.}\ \bibnamefont {Zhang}}, \bibinfo {author} {\bibfnamefont {Z.-D.}\ \bibnamefont {Li}}, \bibinfo {author} {\bibfnamefont {R.}~\bibnamefont {Zhang}}, \bibinfo {author} {\bibfnamefont {X.-F.}\ \bibnamefont {Yin}}, \bibinfo {author} {\bibfnamefont {Y.-Y.}\ \bibnamefont {Fei}}, \bibinfo {author} {\bibfnamefont {L.}~\bibnamefont {Li}}, \bibinfo {author} {\bibfnamefont {N.-L.}\ \bibnamefont {Liu}}, \bibinfo {author} {\bibfnamefont {F.}~\bibnamefont {Xu}}, \bibinfo {author} {\bibfnamefont {Y.-A.}\ \bibnamefont {Chen}},\ and\ \bibinfo {author} {\bibfnamefont {J.-W.}\ \bibnamefont {Pan}},\ }\bibfield  {title} {\bibinfo {title} {Distributed quantum phase estimation with entangled photons},\ }\href {https://doi.org/https://doi.org/10.1038/s41566-020-00718-2} {\bibfield  {journal} {\bibinfo  {journal} {Nat. Photonics}\ }\textbf {\bibinfo {volume} {15}},\ \bibinfo {pages} {137} (\bibinfo {year}
  {2021})}\BibitemShut {NoStop}%
\bibitem [{\citenamefont {Malia}\ \emph {et~al.}(2022)\citenamefont {Malia}, \citenamefont {Wu}, \citenamefont {{Mart{\'i}nez-Rinc{\'o}n}},\ and\ \citenamefont {Kasevich}}]{Malia2022Distributed}%
  \BibitemOpen
  \bibfield  {author} {\bibinfo {author} {\bibfnamefont {B.~K.}\ \bibnamefont {Malia}}, \bibinfo {author} {\bibfnamefont {Y.}~\bibnamefont {Wu}}, \bibinfo {author} {\bibfnamefont {J.}~\bibnamefont {{Mart{\'i}nez-Rinc{\'o}n}}},\ and\ \bibinfo {author} {\bibfnamefont {M.~A.}\ \bibnamefont {Kasevich}},\ }\bibfield  {title} {\bibinfo {title} {Distributed quantum sensing with mode-entangled spin-squeezed atomic states},\ }\href {https://doi.org/10.1038/s41586-022-05363-z} {\bibfield  {journal} {\bibinfo  {journal} {Nature}\ }\textbf {\bibinfo {volume} {612}},\ \bibinfo {pages} {661} (\bibinfo {year} {2022})}\BibitemShut {NoStop}%
\bibitem [{\citenamefont {Zhang}\ and\ \citenamefont {Zhuang}(2021)}]{Zhuang_2021}%
  \BibitemOpen
  \bibfield  {author} {\bibinfo {author} {\bibfnamefont {Z.}~\bibnamefont {Zhang}}\ and\ \bibinfo {author} {\bibfnamefont {Q.}~\bibnamefont {Zhuang}},\ }\bibfield  {title} {\bibinfo {title} {Distributed quantum sensing},\ }\href {https://doi.org/10.1088/2058-9565/abd4c3} {\bibfield  {journal} {\bibinfo  {journal} {Quantum Sci. Technol.}\ }\textbf {\bibinfo {volume} {6}},\ \bibinfo {pages} {043001} (\bibinfo {year} {2021})}\BibitemShut {NoStop}%
\bibitem [{\citenamefont {Yang}\ \emph {et~al.}(2024)\citenamefont {Yang}, \citenamefont {Yadin},\ and\ \citenamefont {Xu}}]{Network_XuPRL_2024}%
  \BibitemOpen
  \bibfield  {author} {\bibinfo {author} {\bibfnamefont {Y.}~\bibnamefont {Yang}}, \bibinfo {author} {\bibfnamefont {B.}~\bibnamefont {Yadin}},\ and\ \bibinfo {author} {\bibfnamefont {Z.-P.}\ \bibnamefont {Xu}},\ }\bibfield  {title} {\bibinfo {title} {{Quantum-Enhanced Metrology} with {Network States}},\ }\href {https://link.aps.org/doi/10.1103/PhysRevLett.132.210801} {\bibfield  {journal} {\bibinfo  {journal} {Phys. Rev. Lett.}\ }\textbf {\bibinfo {volume} {132}},\ \bibinfo {pages} {210801} (\bibinfo {year} {2024})}\BibitemShut {NoStop}%
\bibitem [{\citenamefont {Gottesman}\ \emph {et~al.}(2012)\citenamefont {Gottesman}, \citenamefont {Jennewein},\ and\ \citenamefont {Croke}}]{LongBaseLine2012}%
  \BibitemOpen
  \bibfield  {author} {\bibinfo {author} {\bibfnamefont {D.}~\bibnamefont {Gottesman}}, \bibinfo {author} {\bibfnamefont {T.}~\bibnamefont {Jennewein}},\ and\ \bibinfo {author} {\bibfnamefont {S.}~\bibnamefont {Croke}},\ }\bibfield  {title} {\bibinfo {title} {{Longer-Baseline Telescopes Using Quantum Repeaters}},\ }\href {https://doi.org/10.1103/PhysRevLett.109.070503} {\bibfield  {journal} {\bibinfo  {journal} {Phys. Rev. Lett.}\ }\textbf {\bibinfo {volume} {109}},\ \bibinfo {pages} {070503} (\bibinfo {year} {2012})}\BibitemShut {NoStop}%
\bibitem [{\citenamefont {Stas}\ \emph {et~al.}(2026)\citenamefont {Stas}, \citenamefont {Wei}, \citenamefont {Sirotin}, \citenamefont {Huan}, \citenamefont {Yazlar}, \citenamefont {Abdo~Arias}, \citenamefont {Knyazev}, \citenamefont {Baranes}, \citenamefont {Machielse}, \citenamefont {Grandi}, \citenamefont {Riedel}, \citenamefont {Borregaard}, \citenamefont {Park}, \citenamefont {Lončar}, \citenamefont {Suleymanzade},\ and\ \citenamefont {Lukin}}]{Stas2026Entanglementassisted}%
  \BibitemOpen
  \bibfield  {author} {\bibinfo {author} {\bibfnamefont {P.-J.}\ \bibnamefont {Stas}}, \bibinfo {author} {\bibfnamefont {Y.-C.}\ \bibnamefont {Wei}}, \bibinfo {author} {\bibfnamefont {M.}~\bibnamefont {Sirotin}}, \bibinfo {author} {\bibfnamefont {Y.~Q.}\ \bibnamefont {Huan}}, \bibinfo {author} {\bibfnamefont {U.}~\bibnamefont {Yazlar}}, \bibinfo {author} {\bibfnamefont {F.}~\bibnamefont {Abdo~Arias}}, \bibinfo {author} {\bibfnamefont {E.}~\bibnamefont {Knyazev}}, \bibinfo {author} {\bibfnamefont {G.}~\bibnamefont {Baranes}}, \bibinfo {author} {\bibfnamefont {B.}~\bibnamefont {Machielse}}, \bibinfo {author} {\bibfnamefont {S.}~\bibnamefont {Grandi}}, \bibinfo {author} {\bibfnamefont {D.}~\bibnamefont {Riedel}}, \bibinfo {author} {\bibfnamefont {J.}~\bibnamefont {Borregaard}}, \bibinfo {author} {\bibfnamefont {H.}~\bibnamefont {Park}}, \bibinfo {author} {\bibfnamefont {M.}~\bibnamefont {Lončar}}, \bibinfo {author} {\bibfnamefont {A.}~\bibnamefont {Suleymanzade}},\ and\ \bibinfo {author} {\bibfnamefont {M.~D.}\
  \bibnamefont {Lukin}},\ }\bibfield  {title} {\bibinfo {title} {Entanglement-assisted non-local optical interferometry in a quantum network},\ }\href {https://doi.org/10.1038/s41586-026-10171-w} {\bibfield  {journal} {\bibinfo  {journal} {Nature}\ }\textbf {\bibinfo {volume} {651}},\ \bibinfo {pages} {326} (\bibinfo {year} {2026})}\BibitemShut {NoStop}%
\bibitem [{\citenamefont {Wang}\ \emph {et~al.}(2026)\citenamefont {Wang}, \citenamefont {Luo}, \citenamefont {Gao}, \citenamefont {Liu}, \citenamefont {Wang}, \citenamefont {Yan}, \citenamefont {Ke}, \citenamefont {Teng}, \citenamefont {Zheng}, \citenamefont {Cao}, \citenamefont {Li}, \citenamefont {Peng}, \citenamefont {Zhang}, \citenamefont {Bao},\ and\ \citenamefont {Pan}}]{Wang2026Memoryassisted}%
  \BibitemOpen
  \bibfield  {author} {\bibinfo {author} {\bibfnamefont {B.}~\bibnamefont {Wang}}, \bibinfo {author} {\bibfnamefont {X.-Y.}\ \bibnamefont {Luo}}, \bibinfo {author} {\bibfnamefont {B.-F.}\ \bibnamefont {Gao}}, \bibinfo {author} {\bibfnamefont {J.-L.}\ \bibnamefont {Liu}}, \bibinfo {author} {\bibfnamefont {C.-Y.}\ \bibnamefont {Wang}}, \bibinfo {author} {\bibfnamefont {Z.}~\bibnamefont {Yan}}, \bibinfo {author} {\bibfnamefont {Q.-M.}\ \bibnamefont {Ke}}, \bibinfo {author} {\bibfnamefont {D.}~\bibnamefont {Teng}}, \bibinfo {author} {\bibfnamefont {M.-Y.}\ \bibnamefont {Zheng}}, \bibinfo {author} {\bibfnamefont {Y.}~\bibnamefont {Cao}}, \bibinfo {author} {\bibfnamefont {J.}~\bibnamefont {Li}}, \bibinfo {author} {\bibfnamefont {C.-Z.}\ \bibnamefont {Peng}}, \bibinfo {author} {\bibfnamefont {Q.}~\bibnamefont {Zhang}}, \bibinfo {author} {\bibfnamefont {X.-H.}\ \bibnamefont {Bao}},\ and\ \bibinfo {author} {\bibfnamefont {J.-W.}\ \bibnamefont {Pan}},\ }\bibfield  {title} {\bibinfo {title} {Memory-assisted nonlocal
  interferometer toward long-baseline telescopes},\ }\href {https://doi.org/10.1103/qpzn-h7p9} {\bibfield  {journal} {\bibinfo  {journal} {Phys. Rev. Lett.}\ }\textbf {\bibinfo {volume} {136}},\ \bibinfo {pages} {240801} (\bibinfo {year} {2026})}\BibitemShut {NoStop}%
\bibitem [{\citenamefont {Wang}\ \emph {et~al.}(2023)\citenamefont {Wang}, \citenamefont {Wang}, \citenamefont {Zuo}, \citenamefont {Dong}, \citenamefont {Siviloglou},\ and\ \citenamefont {Chen}}]{Wang2023Thermometry}%
  \BibitemOpen
  \bibfield  {author} {\bibinfo {author} {\bibfnamefont {X.}~\bibnamefont {Wang}}, \bibinfo {author} {\bibfnamefont {J.}~\bibnamefont {Wang}}, \bibinfo {author} {\bibfnamefont {Y.}~\bibnamefont {Zuo}}, \bibinfo {author} {\bibfnamefont {L.}~\bibnamefont {Dong}}, \bibinfo {author} {\bibfnamefont {G.~A.}\ \bibnamefont {Siviloglou}},\ and\ \bibinfo {author} {\bibfnamefont {J.}~\bibnamefont {Chen}},\ }\bibfield  {title} {\bibinfo {title} {Thermometry utilizing stored short-wavelength spin waves in cold atomic ensembles},\ }\href {https://doi.org/10.1088/1674-1056/accb4f} {\bibfield  {journal} {\bibinfo  {journal} {Chin. Phys. B}\ }\textbf {\bibinfo {volume} {32}},\ \bibinfo {pages} {074206} (\bibinfo {year} {2023})}\BibitemShut {NoStop}%
\bibitem [{\citenamefont {Steck}()}]{SteckRb85DLine}%
  \BibitemOpen
  \bibfield  {author} {\bibinfo {author} {\bibfnamefont {D.~A.}\ \bibnamefont {Steck}},\ }\href@noop {} {\bibinfo {title} {Rubidium 85 {D Line Data}}},\ \bibinfo {note} {available online at http://steck.us/alkalidata (revision 2.3.4, 8 August 2025)}\BibitemShut {NoStop}%
\bibitem [{\citenamefont {Novikova}\ \emph {et~al.}(2007)\citenamefont {Novikova}, \citenamefont {Gorshkov}, \citenamefont {Phillips}, \citenamefont {S{\o}rensen}, \citenamefont {Lukin},\ and\ \citenamefont {Walsworth}}]{Novikova2007Optimal}%
  \BibitemOpen
  \bibfield  {author} {\bibinfo {author} {\bibfnamefont {I.}~\bibnamefont {Novikova}}, \bibinfo {author} {\bibfnamefont {A.~V.}\ \bibnamefont {Gorshkov}}, \bibinfo {author} {\bibfnamefont {D.~F.}\ \bibnamefont {Phillips}}, \bibinfo {author} {\bibfnamefont {A.~S.}\ \bibnamefont {S{\o}rensen}}, \bibinfo {author} {\bibfnamefont {M.~D.}\ \bibnamefont {Lukin}},\ and\ \bibinfo {author} {\bibfnamefont {R.~L.}\ \bibnamefont {Walsworth}},\ }\bibfield  {title} {\bibinfo {title} {Optimal {{Control}} of {{Light Pulse Storage}} and {{Retrieval}}},\ }\href {https://link.aps.org/doi/10.1103/PhysRevLett.98.243602} {\bibfield  {journal} {\bibinfo  {journal} {Phys. Rev. Lett.}\ }\textbf {\bibinfo {volume} {98}} (\bibinfo {year} {2007})}\BibitemShut {NoStop}%
\bibitem [{\citenamefont {Reim}\ \emph {et~al.}(2012)\citenamefont {Reim}, \citenamefont {Nunn}, \citenamefont {Jin}, \citenamefont {Michelberger}, \citenamefont {Champion}, \citenamefont {England}, \citenamefont {Lee}, \citenamefont {Kolthammer}, \citenamefont {Langford},\ and\ \citenamefont {Walmsley}}]{Reim2012Multipulse}%
  \BibitemOpen
  \bibfield  {author} {\bibinfo {author} {\bibfnamefont {K.~F.}\ \bibnamefont {Reim}}, \bibinfo {author} {\bibfnamefont {J.}~\bibnamefont {Nunn}}, \bibinfo {author} {\bibfnamefont {X.-M.}\ \bibnamefont {Jin}}, \bibinfo {author} {\bibfnamefont {P.~S.}\ \bibnamefont {Michelberger}}, \bibinfo {author} {\bibfnamefont {T.~F.~M.}\ \bibnamefont {Champion}}, \bibinfo {author} {\bibfnamefont {D.~G.}\ \bibnamefont {England}}, \bibinfo {author} {\bibfnamefont {K.~C.}\ \bibnamefont {Lee}}, \bibinfo {author} {\bibfnamefont {W.~S.}\ \bibnamefont {Kolthammer}}, \bibinfo {author} {\bibfnamefont {N.~K.}\ \bibnamefont {Langford}},\ and\ \bibinfo {author} {\bibfnamefont {I.~A.}\ \bibnamefont {Walmsley}},\ }\bibfield  {title} {\bibinfo {title} {Multipulse {Addressing} of a {Raman} {Quantum} {Memory}: {Configurable} {Beam} {Splitting} and {Efficient} {Readout}},\ }\href {https://doi.org/10.1103/PhysRevLett.108.263602} {\bibfield  {journal} {\bibinfo  {journal} {Phys. Rev. Lett.}\ }\textbf {\bibinfo {volume} {108}},\ \bibinfo
  {pages} {263602} (\bibinfo {year} {2012})}\BibitemShut {NoStop}%
\bibitem [{\citenamefont {Wang}\ \emph {et~al.}(2024)\citenamefont {Wang}, \citenamefont {Dong}, \citenamefont {Wang}, \citenamefont {Zhou}, \citenamefont {Huang}, \citenamefont {Zuo}, \citenamefont {Siviloglou},\ and\ \citenamefont {Chen}}]{Wang_PhysRevResearch2024}%
  \BibitemOpen
  \bibfield  {author} {\bibinfo {author} {\bibfnamefont {J.}~\bibnamefont {Wang}}, \bibinfo {author} {\bibfnamefont {L.}~\bibnamefont {Dong}}, \bibinfo {author} {\bibfnamefont {X.}~\bibnamefont {Wang}}, \bibinfo {author} {\bibfnamefont {Z.}~\bibnamefont {Zhou}}, \bibinfo {author} {\bibfnamefont {J.}~\bibnamefont {Huang}}, \bibinfo {author} {\bibfnamefont {Y.}~\bibnamefont {Zuo}}, \bibinfo {author} {\bibfnamefont {G.~A.}\ \bibnamefont {Siviloglou}},\ and\ \bibinfo {author} {\bibfnamefont {J.~F.}\ \bibnamefont {Chen}},\ }\bibfield  {title} {\bibinfo {title} {Light-induced fictitious magnetic fields for quantum storage in cold atomic ensembles},\ }\href {https://doi.org/10.1103/PhysRevResearch.6.L042002} {\bibfield  {journal} {\bibinfo  {journal} {Phys. Rev. Res.}\ }\textbf {\bibinfo {volume} {6}},\ \bibinfo {pages} {L042002} (\bibinfo {year} {2024})}\BibitemShut {NoStop}%
\bibitem [{\citenamefont {Cohen}\ \emph {et~al.}(2019)\citenamefont {Cohen}, \citenamefont {Jadeja}, \citenamefont {Sula}, \citenamefont {Venturelli}, \citenamefont {Deans}, \citenamefont {Marmugi},\ and\ \citenamefont {Renzoni}}]{Cohen2019Cold}%
  \BibitemOpen
  \bibfield  {author} {\bibinfo {author} {\bibfnamefont {Y.}~\bibnamefont {Cohen}}, \bibinfo {author} {\bibfnamefont {K.}~\bibnamefont {Jadeja}}, \bibinfo {author} {\bibfnamefont {S.}~\bibnamefont {Sula}}, \bibinfo {author} {\bibfnamefont {M.}~\bibnamefont {Venturelli}}, \bibinfo {author} {\bibfnamefont {C.}~\bibnamefont {Deans}}, \bibinfo {author} {\bibfnamefont {L.}~\bibnamefont {Marmugi}},\ and\ \bibinfo {author} {\bibfnamefont {F.}~\bibnamefont {Renzoni}},\ }\bibfield  {title} {\bibinfo {title} {A cold atom radio-frequency magnetometer},\ }\href {https://doi.org/10.1063/1.5084004} {\bibfield  {journal} {\bibinfo  {journal} {Appl. Phys. Lett.}\ }\textbf {\bibinfo {volume} {114}},\ \bibinfo {pages} {073505} (\bibinfo {year} {2019})}\BibitemShut {NoStop}%
\bibitem [{\citenamefont {Cassens}\ \emph {et~al.}(2025)\citenamefont {Cassens}, \citenamefont {Meyer-Hoppe}, \citenamefont {Rasel},\ and\ \citenamefont {Klempt}}]{cassens2025entanglement}%
  \BibitemOpen
  \bibfield  {author} {\bibinfo {author} {\bibfnamefont {C.}~\bibnamefont {Cassens}}, \bibinfo {author} {\bibfnamefont {B.}~\bibnamefont {Meyer-Hoppe}}, \bibinfo {author} {\bibfnamefont {E.}~\bibnamefont {Rasel}},\ and\ \bibinfo {author} {\bibfnamefont {C.}~\bibnamefont {Klempt}},\ }\bibfield  {title} {\bibinfo {title} {Entanglement-enhanced atomic gravimeter},\ }\href {https://doi.org/10.1103/PhysRevX.15.011029} {\bibfield  {journal} {\bibinfo  {journal} {Phys. Rev. X}\ }\textbf {\bibinfo {volume} {15}},\ \bibinfo {pages} {011029} (\bibinfo {year} {2025})}\BibitemShut {NoStop}%
\bibitem [{\citenamefont {Liu}\ \emph {et~al.}(2022)\citenamefont {Liu}, \citenamefont {Wu}, \citenamefont {Cao}, \citenamefont {Mao}, \citenamefont {Li}, \citenamefont {Guo}, \citenamefont {Tey},\ and\ \citenamefont {You}}]{liu2022nonlinear}%
  \BibitemOpen
  \bibfield  {author} {\bibinfo {author} {\bibfnamefont {Q.}~\bibnamefont {Liu}}, \bibinfo {author} {\bibfnamefont {L.-N.}\ \bibnamefont {Wu}}, \bibinfo {author} {\bibfnamefont {J.-H.}\ \bibnamefont {Cao}}, \bibinfo {author} {\bibfnamefont {T.-W.}\ \bibnamefont {Mao}}, \bibinfo {author} {\bibfnamefont {X.-W.}\ \bibnamefont {Li}}, \bibinfo {author} {\bibfnamefont {S.-F.}\ \bibnamefont {Guo}}, \bibinfo {author} {\bibfnamefont {M.~K.}\ \bibnamefont {Tey}},\ and\ \bibinfo {author} {\bibfnamefont {L.}~\bibnamefont {You}},\ }\bibfield  {title} {\bibinfo {title} {Nonlinear interferometry beyond classical limit enabled by cyclic dynamics},\ }\href {https://doi.org/10.1038/s41567-021-01441-7} {\bibfield  {journal} {\bibinfo  {journal} {Nat. Phys.}\ }\textbf {\bibinfo {volume} {18}},\ \bibinfo {pages} {167} (\bibinfo {year} {2022})}\BibitemShut {NoStop}%
\bibitem [{\citenamefont {Yan}\ \emph {et~al.}(2017)\citenamefont {Yan}, \citenamefont {Wu}, \citenamefont {Jia}, \citenamefont {Liu}, \citenamefont {Deng}, \citenamefont {Li}, \citenamefont {Wang}, \citenamefont {Xie},\ and\ \citenamefont {Peng}}]{Yan2017Establishing}%
  \BibitemOpen
  \bibfield  {author} {\bibinfo {author} {\bibfnamefont {Z.}~\bibnamefont {Yan}}, \bibinfo {author} {\bibfnamefont {L.}~\bibnamefont {Wu}}, \bibinfo {author} {\bibfnamefont {X.}~\bibnamefont {Jia}}, \bibinfo {author} {\bibfnamefont {Y.}~\bibnamefont {Liu}}, \bibinfo {author} {\bibfnamefont {R.}~\bibnamefont {Deng}}, \bibinfo {author} {\bibfnamefont {S.}~\bibnamefont {Li}}, \bibinfo {author} {\bibfnamefont {H.}~\bibnamefont {Wang}}, \bibinfo {author} {\bibfnamefont {C.}~\bibnamefont {Xie}},\ and\ \bibinfo {author} {\bibfnamefont {K.}~\bibnamefont {Peng}},\ }\bibfield  {title} {\bibinfo {title} {Establishing and storing of deterministic quantum entanglement among three distant atomic ensembles},\ }\href {https://doi.org/10.1038/s41467-017-00809-9} {\bibfield  {journal} {\bibinfo  {journal} {Nat. Commun.}\ }\textbf {\bibinfo {volume} {8}},\ \bibinfo {pages} {718} (\bibinfo {year} {2017})}\BibitemShut {NoStop}%
\bibitem [{\citenamefont {Monnier}(2003)}]{John_D_Monnier_2003}%
  \BibitemOpen
  \bibfield  {author} {\bibinfo {author} {\bibfnamefont {J.~D.}\ \bibnamefont {Monnier}},\ }\bibfield  {title} {\bibinfo {title} {Optical interferometry in astronomy},\ }\href {https://doi.org/10.1088/0034-4885/66/5/203} {\bibfield  {journal} {\bibinfo  {journal} {Rep. Prog. Phys.}\ }\textbf {\bibinfo {volume} {66}},\ \bibinfo {pages} {789} (\bibinfo {year} {2003})}\BibitemShut {NoStop}%
\bibitem [{\citenamefont {Guo}\ \emph {et~al.}(2020{\natexlab{b}})\citenamefont {Guo}, \citenamefont {Breum}, \citenamefont {Borregaard}, \citenamefont {Izumi}, \citenamefont {Larsen}, \citenamefont {Gehring}, \citenamefont {Christandl}, \citenamefont {Neergaard-Nielsen},\ and\ \citenamefont {Andersen}}]{Guo2020Distributed}%
  \BibitemOpen
  \bibfield  {author} {\bibinfo {author} {\bibfnamefont {X.}~\bibnamefont {Guo}}, \bibinfo {author} {\bibfnamefont {C.~R.}\ \bibnamefont {Breum}}, \bibinfo {author} {\bibfnamefont {J.}~\bibnamefont {Borregaard}}, \bibinfo {author} {\bibfnamefont {S.}~\bibnamefont {Izumi}}, \bibinfo {author} {\bibfnamefont {M.~V.}\ \bibnamefont {Larsen}}, \bibinfo {author} {\bibfnamefont {T.}~\bibnamefont {Gehring}}, \bibinfo {author} {\bibfnamefont {M.}~\bibnamefont {Christandl}}, \bibinfo {author} {\bibfnamefont {J.~S.}\ \bibnamefont {Neergaard-Nielsen}},\ and\ \bibinfo {author} {\bibfnamefont {U.~L.}\ \bibnamefont {Andersen}},\ }\bibfield  {title} {\bibinfo {title} {Distributed quantum sensing in a continuous-variable entangled network},\ }\href {https://doi.org/10.1038/s41567-019-0743-x} {\bibfield  {journal} {\bibinfo  {journal} {Nature Physics}\ }\textbf {\bibinfo {volume} {16}},\ \bibinfo {pages} {281} (\bibinfo {year} {2020}{\natexlab{b}})}\BibitemShut {NoStop}%
\bibitem [{\citenamefont {Xia}\ \emph {et~al.}(2020)\citenamefont {Xia}, \citenamefont {Li}, \citenamefont {Clark}, \citenamefont {Hart}, \citenamefont {Zhuang},\ and\ \citenamefont {Zhang}}]{Xia2020Demonstration}%
  \BibitemOpen
  \bibfield  {author} {\bibinfo {author} {\bibfnamefont {Y.}~\bibnamefont {Xia}}, \bibinfo {author} {\bibfnamefont {W.}~\bibnamefont {Li}}, \bibinfo {author} {\bibfnamefont {W.}~\bibnamefont {Clark}}, \bibinfo {author} {\bibfnamefont {D.}~\bibnamefont {Hart}}, \bibinfo {author} {\bibfnamefont {Q.}~\bibnamefont {Zhuang}},\ and\ \bibinfo {author} {\bibfnamefont {Z.}~\bibnamefont {Zhang}},\ }\bibfield  {title} {\bibinfo {title} {Demonstration of a reconfigurable entangled radio-frequency photonic sensor network},\ }\href {https://doi.org/10.1103/PhysRevLett.124.150502} {\bibfield  {journal} {\bibinfo  {journal} {Phys. Rev. Lett.}\ }\textbf {\bibinfo {volume} {124}},\ \bibinfo {pages} {150502} (\bibinfo {year} {2020})}\BibitemShut {NoStop}%
\bibitem [{\citenamefont {Yuen}\ and\ \citenamefont {Chan}(1983)}]{Yuen1983Noise}%
  \BibitemOpen
  \bibfield  {author} {\bibinfo {author} {\bibfnamefont {H.~P.}\ \bibnamefont {Yuen}}\ and\ \bibinfo {author} {\bibfnamefont {V.~W.~S.}\ \bibnamefont {Chan}},\ }\bibfield  {title} {\bibinfo {title} {Noise in homodyne and heterodyne detection},\ }\href {https://doi.org/10.1364/OL.8.000177} {\bibfield  {journal} {\bibinfo  {journal} {Opt. Lett.}\ }\textbf {\bibinfo {volume} {8}},\ \bibinfo {pages} {177} (\bibinfo {year} {1983})}\BibitemShut {NoStop}%
\bibitem [{\citenamefont {Buonanno}\ \emph {et~al.}(2003)\citenamefont {Buonanno}, \citenamefont {Chen},\ and\ \citenamefont {Mavalvala}}]{Buonanno2003Quantum}%
  \BibitemOpen
  \bibfield  {author} {\bibinfo {author} {\bibfnamefont {A.}~\bibnamefont {Buonanno}}, \bibinfo {author} {\bibfnamefont {Y.}~\bibnamefont {Chen}},\ and\ \bibinfo {author} {\bibfnamefont {N.}~\bibnamefont {Mavalvala}},\ }\bibfield  {title} {\bibinfo {title} {Quantum noise in laser-interferometer gravitational-wave detectors with a heterodyne readout scheme},\ }\href {https://doi.org/10.1103/PhysRevD.67.122005} {\bibfield  {journal} {\bibinfo  {journal} {Phys. Rev. D}\ }\textbf {\bibinfo {volume} {67}},\ \bibinfo {pages} {122005} (\bibinfo {year} {2003})}\BibitemShut {NoStop}%
\bibitem [{\citenamefont {Chen}\ \emph {et~al.}(2025)\citenamefont {Chen}, \citenamefont {Lin}, \citenamefont {Shiu}, \citenamefont {Zhong}, \citenamefont {Tseng},\ and\ \citenamefont {Chen}}]{Chen2025Highefficiency}%
  \BibitemOpen
  \bibfield  {author} {\bibinfo {author} {\bibfnamefont {L.-C.}\ \bibnamefont {Chen}}, \bibinfo {author} {\bibfnamefont {M.-Y.}\ \bibnamefont {Lin}}, \bibinfo {author} {\bibfnamefont {J.-S.}\ \bibnamefont {Shiu}}, \bibinfo {author} {\bibfnamefont {X.-Q.}\ \bibnamefont {Zhong}}, \bibinfo {author} {\bibfnamefont {P.-H.}\ \bibnamefont {Tseng}},\ and\ \bibinfo {author} {\bibfnamefont {Y.-F.}\ \bibnamefont {Chen}},\ }\bibfield  {title} {\bibinfo {title} {High-efficiency telecom frequency conversion via a diamond-type atomic ensemble},\ }\href {https://doi.org/10.1103/mp7d-87nw} {\bibfield  {journal} {\bibinfo  {journal} {Phys. Rev. A}\ }\textbf {\bibinfo {volume} {112}},\ \bibinfo {pages} {013709} (\bibinfo {year} {2025})}\BibitemShut {NoStop}%
\bibitem [{\citenamefont {Luo}\ \emph {et~al.}(2026)\citenamefont {Luo}, \citenamefont {Wang}, \citenamefont {Zheng}, \citenamefont {Wang}, \citenamefont {Liu}, \citenamefont {Gao}, \citenamefont {Li}, \citenamefont {Yan}, \citenamefont {Ke}, \citenamefont {Teng}, \citenamefont {Wang}, \citenamefont {Wu}, \citenamefont {Huang}, \citenamefont {Li}, \citenamefont {You}, \citenamefont {Xie}, \citenamefont {Xu}, \citenamefont {Zhang}, \citenamefont {Bao},\ and\ \citenamefont {Pan}}]{Luo2025Entangling}%
  \BibitemOpen
  \bibfield  {author} {\bibinfo {author} {\bibfnamefont {X.-Y.}\ \bibnamefont {Luo}}, \bibinfo {author} {\bibfnamefont {C.-Y.}\ \bibnamefont {Wang}}, \bibinfo {author} {\bibfnamefont {M.-Y.}\ \bibnamefont {Zheng}}, \bibinfo {author} {\bibfnamefont {B.}~\bibnamefont {Wang}}, \bibinfo {author} {\bibfnamefont {J.-L.}\ \bibnamefont {Liu}}, \bibinfo {author} {\bibfnamefont {B.-F.}\ \bibnamefont {Gao}}, \bibinfo {author} {\bibfnamefont {J.}~\bibnamefont {Li}}, \bibinfo {author} {\bibfnamefont {Z.}~\bibnamefont {Yan}}, \bibinfo {author} {\bibfnamefont {Q.-M.}\ \bibnamefont {Ke}}, \bibinfo {author} {\bibfnamefont {D.}~\bibnamefont {Teng}}, \bibinfo {author} {\bibfnamefont {R.-C.}\ \bibnamefont {Wang}}, \bibinfo {author} {\bibfnamefont {J.}~\bibnamefont {Wu}}, \bibinfo {author} {\bibfnamefont {J.}~\bibnamefont {Huang}}, \bibinfo {author} {\bibfnamefont {H.}~\bibnamefont {Li}}, \bibinfo {author} {\bibfnamefont {L.-X.}\ \bibnamefont {You}}, \bibinfo {author} {\bibfnamefont {X.-P.}\ \bibnamefont {Xie}}, \bibinfo {author}
  {\bibfnamefont {F.}~\bibnamefont {Xu}}, \bibinfo {author} {\bibfnamefont {Q.}~\bibnamefont {Zhang}}, \bibinfo {author} {\bibfnamefont {X.-H.}\ \bibnamefont {Bao}},\ and\ \bibinfo {author} {\bibfnamefont {J.-W.}\ \bibnamefont {Pan}},\ }\bibfield  {title} {\bibinfo {title} {Entangling quantum memories through a 420 km long fiber},\ }\href {https://doi.org/10.1103/ccd6-rf1s} {\bibfield  {journal} {\bibinfo  {journal} {Phys. Rev. Lett.}\ }\textbf {\bibinfo {volume} {137}},\ \bibinfo {pages} {070801} (\bibinfo {year} {2026})}\BibitemShut {NoStop}%
\end{thebibliography}%

\clearpage

\begin{appendix}\label{appendix}

\section{Experimental setup details} 

In the experiment illustrated in Fig.~\ref{fig:setup}(a), an elongated laser-cooled $^{85}\text{Rb}$ atomic ensemble loaded in a two-dimensional MOT acts as a beam splitter based on electromagnetically-induced transparency optical storage. Initially, the atoms are trapped and cooled for $\SI{18.7}{ms}$. After MOT loading, we initialize all atoms into the ground state $|g:5^{2}S_{1/2},F=2,m_{F}=2\rangle$ using $\SI{0.3}{ms}$ of optical pumping. The $\sigma^{+}$ signal light $E$, resonant with the transition $|g\rangle\rightarrow|e:5^{2}P_{1/2},F=3,m_{F}=3\rangle$, passes through the atomic medium, which becomes transparent in the presence of a strong $\sigma^{+}$ control light $\Omega_{c}$. This control beam is resonant with the optical transition $|e\rangle\rightarrow|s:5^{2}S_{1/2},F=3,m_{F}=2\rangle$. When the control light is switched off, the signal light is partially transmitted and is considered the transmission of the beam splitter, while the absorbed light is converted to a long-lived atomic collective excitation that leads to reflection. After a storage time of $\Delta\tau$, this stored excitation accumulates a phase $\Delta\phi$ induced by the magnetic field ${B}_z$ along the $z$-direction, which coincides with the long axis of the atomic cloud. A second control pulse reads out the stored atomic excitation, converting it back into the optical mode that constitutes the retrieved signal. This memory process is repeated $N$ times before the next round of MOT loading. Each full cycle, including MOT loading and the write-read process, lasts $\SI{20}{ms}$, within which the measurement window accounts for $\SI{1}{ms}$. \par

\section{Visibility of direct current addition} 

Taking the response function for the homodyne detector as $k(t)$, the output current is:
\begin{equation}
    \begin{aligned}
        i_{+}(t)&=i_{T}(t)+i_{R}(t) \\
        &\propto |\mathcal{E}|\int d\tau k(t-\tau)[X_{T}(\tau)+X_{R}(\tau)], \\
    \end{aligned}
\end{equation}
where $X_{T}(\tau)=E_{T}a_{T}(\tau)e^{-i\delta\omega\tau}e^{i\delta\phi_{T}}+c.c.$ and $X_{R}(\tau)=E_{R}a_{R}(\tau-\Delta\tau)e^{-i\delta\omega(\tau-\Delta\tau)}e^{i\delta\phi_{R}}+c.c.$, with $a_{j}(\tau)$, $(j=T,R)$, being the normalized pulse profiles satisfying $\int d\tau a_{j}^{2}(\tau)=1$. All four correlation functions for the $X_{T}$ and $X_{R}$ are: 
\begin{subequations}
\allowdisplaybreaks
    \begin{align}
        &\langle X_{T}(\tau)X_{T}(\tau^{\prime})\rangle \notag\\
        =&\langle E_{T}a_{T}(\tau)E_{T}^{*}a_{T}(\tau^{\prime})\rangle e^{-i\delta\omega(\tau-\tau^{\prime})}+c. c. \notag\\
        =&I_{T}\mathcal{A}_{TT}(\tau-\tau^{\prime})e^{-i\delta\omega(\tau-\tau^{\prime})}+c. c.,\\
        &\langle X_{R}(\tau)X_{R}(\tau^{\prime})\rangle \notag\\
        =&\langle E_{R}a_{R}(\tau-\Delta\tau)E_{R}^{*}a_{R}(\tau^{\prime}-\Delta\tau)\rangle e^{-i\delta\omega(\tau-\tau^{\prime})}+c. c. \notag\\
        =&I_{R}\mathcal{A}_{RR}(\tau-\tau^{\prime})e^{-i\delta\omega(\tau-\tau^{\prime})}+c. c., \\
        &\langle X_{T}(\tau)X_{R}(\tau^{\prime})\rangle \notag\\
        =&\langle E_{T}a_{T}(\tau)E_{R}^{*}a_{R}(\tau^{\prime}-\Delta\tau)\rangle e^{-i\delta\omega(\tau-(\tau^{\prime}-\Delta\tau))}e^{i\Delta\phi}+c. c. \notag\\
        =&\sqrt{I_{T}I_{R}}\mathcal{A}_{TR}(\tau-\tau^{\prime}+\Delta\tau)e^{-i\delta\omega(\tau-\tau^{\prime}+\Delta\tau)}e^{i\Delta\phi}+c. c., \\
        &\langle X_{R}(\tau)X_{T}(\tau^{\prime})\rangle \notag\\
        =&\langle E_{R}a_{R}(\tau-\Delta\tau)E_{T}^{*}a_{T}(\tau^{\prime})\rangle e^{-i\delta\omega((\tau-\Delta\tau)-\tau^{\prime})}e^{-i\Delta\phi}+c. c. \notag\\
        =&\sqrt{I_{R}I_{T}}\mathcal{A}_{RT}(\tau-\tau^{\prime}-\Delta\tau)e^{-i\delta\omega(\tau-\tau^{\prime}-\Delta\tau)}e^{-i\Delta\phi}+c. c., 
    \end{align}
\end{subequations}
where $I_{j}=E_{j}E^{*}_{j}$ is the light intensity for $j=T,R$, and $\mathcal{A}_{ij}(\tau)=\langle a_{i}(t)a_{j}(t-\tau)\rangle$ is the corresponding amplitude correlation function. 
The output power is proportional to:
\begin{equation}
    \begin{aligned}
        \langle i_{+}^{2}(t)\rangle= &|\mathcal{E}|^{2}\int d\tau d\tau^{\prime}k(t-\tau)k(t-\tau^{\prime})e^{-i\delta\omega(\tau-\tau^{\prime})}\\
        &\cdot[I_{T}\mathcal{A}_{TT}(\tau-\tau^{\prime})+I_{R}\mathcal{A}_{RR}(\tau-\tau^{\prime})\\
        &+\sqrt{I_{T}I_{R}}\mathcal{A}_{TR}(\tau-\tau^{\prime}+\Delta\tau)e^{-i\delta\omega\Delta\tau}e^{i\Delta\phi}\\
        &+\sqrt{I_{R}I_{T}}\mathcal{A}_{RT}(\tau-\tau^{\prime}-\Delta\tau)e^{i\delta\omega\Delta\tau}e^{-i\Delta\phi}]+c. c. . \\
    \end{aligned}
\end{equation}\par
If the storage time $\Delta\tau$ is significantly larger than the signal pulse width $\Delta T_{c}$, the amplitude correlation functions become $\mathcal{A}_{TR}(\Delta\tau)=\mathcal{A}_{RT}(\Delta\tau)=0$. We use an electronic delay $\Delta T_{e}$ to balance the storage delay; while the phase variation during storage is not compensated, resulting in:
\begin{equation}
\begin{aligned}
	\langle i_{+}^{2}(t)\rangle= &|\mathcal{E}|^{2}\int d\tau d\tau^{\prime}k(t-\tau)k(t-\tau^{\prime})e^{-i\delta\omega(\tau-\tau^{\prime})}\\
        &\cdot[I_{T}\mathcal{A}_{TT}(\tau-\tau^{\prime})+I_{R}\mathcal{A}_{RR}(\tau-\tau^{\prime})\\
        &+\sqrt{I_{T}I_{R}}\mathcal{A}_{TR}(\tau-\tau^{\prime})e^{-i\delta\omega\Delta\tau}e^{i\Delta\phi}\\
        &+\sqrt{I_{R}I_{T}}\mathcal{A}_{RT}(\tau-\tau^{\prime})e^{i\delta\omega\Delta\tau}e^{-i\Delta\phi}]+c. c. . \\
        = &|\mathcal{E}|^{2}\int d\tau^{\prime\prime} \mathcal{K}(\tau^{\prime\prime})e^{-i\delta\omega\tau^{\prime\prime}}\\
        &\cdot[I_{T}\mathcal{A}_{TT}(\tau^{\prime\prime})+I_{R}\mathcal{A}_{RR}(\tau^{\prime\prime})\\
        &+\sqrt{I_{T}I_{R}}\mathcal{A}_{TR}(\tau^{\prime\prime})e^{-i\delta\omega\Delta\tau}e^{i\Delta\phi}\\
        &+\sqrt{I_{R}I_{T}}\mathcal{A}_{RT}(\tau^{\prime\prime})e^{i\delta\omega\Delta\tau}e^{-i\Delta\phi}]+c. c. . \\
\end{aligned}
\end{equation}
where $\mathcal{K}(\tau^{\prime\prime})=\int d\tau k(t-\tau)k(t-\tau+\tau^{\prime\prime})$. \par
The speed of the detector with respect to the integration time can lead to simple analytical expressions for the visibility, $\mathcal{V}$:

(I) If the detector is very slow $k(t-\tau)$ and $k(t-\tau^{\prime})$ can be taken equal to a constant number $k$ over the integration with respect to $\tau$ and $\tau^{\prime}$:
\begin{widetext}
\begin{equation}
    \begin{aligned}
        \langle i_{+}^{2}(t)\rangle =&k^{2}|\mathcal{E}|^{2}\langle \left\{I_{T}\left[\int_{t_{1}}^{t_{2}} d\tau a_{T}(t-\tau)\right]^{2}+I_{R}\left[\int_{t_{1}}^{t_{2}} d\tau a_{R}(t-\tau)\right]^{2}\right.\\
        &+\sqrt{I_{T}I_{R}}e^{-i\delta\omega\Delta\tau}e^{i\Delta\phi}\int_{t_{1}}^{t_{2}} d\tau a_{T}(t-\tau)\int_{t_{1}}^{t_{2}} d\tau a_{R}(t-\tau)\\
        &\left.+\sqrt{I_{R}I_{T}}e^{i\delta\omega\Delta\tau}e^{-i\Delta\phi}\int_{t_{1}}^{t_{2}} d\tau a_{R}(t-\tau)\int_{t_{1}}^{t_{2}} d\tau a_{T}(t-\tau)\right\}+c. c. \rangle\\
        \leq&k^{2}|\mathcal{E}|^{2}(t_{2}-t_{1})\left[I_{T}\langle\int_{t_{1}}^{t_{2}} d\tau a_{T}^{2}(t-\tau)\rangle+I_{R}\langle\int_{t_{1}}^{t_{2}} d\tau a_{R}^{2}(t-\tau)\rangle\right.\\
        &+\sqrt{I_{T}I_{R}}e^{-i\delta\omega\Delta\tau}e^{i\Delta\phi}\langle\sqrt{\int_{t_{1}}^{t_{2}} d\tau a_{T}^{2}(t-\tau)\int_{t_{1}}^{t_{2}} d\tau a_{R}^{2}(t-\tau)}\rangle \\
        &\left.+\sqrt{I_{R}I_{T}}e^{i\delta\omega\Delta\tau}e^{-i\Delta\phi}\langle\sqrt{\int_{t_{1}}^{t_{2}} d\tau a_{R}^{2}(t-\tau)\int_{t_{1}}^{t_{2}} d\tau a_{T}^{2}(t-\tau)}\rangle +c. c. \right]. \\
    \end{aligned}
\end{equation}
\end{widetext}
Here, we use the assumption that the integration region for $d\tau$ and $d\tau^{\prime}$ is from the start time $t_{1}$ to the end time $t_{2}$, then $\int_{t_{1}}^{t_{2}}\int_{t_{1}}^{t_{2}}a(\tau)a(\tau^{\prime}) d\tau d\tau^{\prime}\rightarrow[\int_{t_{1}}^{t_{2}}a(\tau) d\tau]^{2}$, and Cauchy-Schwarz inequality $[\int_{t_{1}}^{t_{2}} a(\tau)d\tau]^{2}\leq(t_{2}-t_{1})\int_{t_{1}}^{t_{2}} d\tau[a(\tau)]^{2}$. Finally, we get Eq. (\ref{eq:iave}) of the main text
\begin{equation}
\langle i^{2}_{+}(t)\rangle\approx2K|\mathcal{E}|^{2}(I_{T}+I_{R})\left[1+\mathcal{V}_{0}\cos{(\delta\omega\Delta\tau-\Delta\phi)}\right], 
\end{equation}
where $K=k^{2}$, and $\mathcal{V}_{0}=2\sqrt{I_{T}I_{R}}/(I_{T}+I_{R})$. \par

(II) If the detector is very fast,  $k(t-\tau)$, $k(t-\tau^{\prime})$ can take the form of a $\delta$-function:
\begin{equation}
    \begin{aligned}
        \langle i_{+}^{2}(t)\rangle \approx&|\mathcal{E}|^{2}[I_{T}\mathcal{A}_{TT}(0)+I_{R}\mathcal{A}_{RR}(0)\\
        &+\sqrt{I_{T}I_{R}}\mathcal{A}_{TR}(0)e^{-i\delta\omega\Delta\tau}e^{i\Delta\phi}\\
        &+\sqrt{I_{R}I_{T}}\mathcal{A}_{RT}(0)e^{i\delta\omega\Delta\tau}e^{-i\Delta\phi}]+c. c. \\
        =&2|\mathcal{E}|^{2}(I_{T}+I_{R})[1+\mathcal{V}\cos{(\delta\omega\Delta\tau-\Delta\phi)}], 
    \end{aligned}
\end{equation}
where $\mathcal{V}=\mathcal{V}_{0}\mathcal{A}_{TR}(0)$. \par

If in any of the two cases above, the self-correlation function can be considered as a constant number. The visibility then only depends on the cross-correlation function and the detector response. If we perform the integration in the frequency domain:
\begin{equation}
    \begin{aligned}
    \mathcal{V}(\delta\omega)&\approx\mathcal{V}_{0}\int d\tau\mathcal{K}(\tau)\mathcal{A}_{TR}(\tau)e^{-i\delta\omega\tau}\\
    &=\mathcal{V}_{0}\int d\omega\tilde{\mathcal{K}}(\omega)\tilde{\mathcal{A}}_{TR}(\delta\omega-\omega),
    \end{aligned}
\end{equation}
where $\tilde{\mathcal{K}}(\omega)$, $\tilde{\mathcal{A}}(\omega)$ are the Fourier transforms of $\mathcal{K}(\tau)$, and $\mathcal{A}(\tau)$ respectively. $\tilde{\mathcal{K}}(\omega) = k^2(\omega)$ is the spectral response of the detector, and $\tilde{\mathcal{A}}(\omega)$ is the pulse spectrum. This is the Eq. (\ref{eq:visibility}) of the main text. \par

\section{Pulse integration and recovery of the interference pattern} 

In the experiment, after every MOT loading, we apply $N$ measurement pulses. The $k$-th pulse starts from the time $t_{1}^{k}$ and ends at $t_{2}^{k}$. The integrated T and R pulses can be written as $i_{T}^{k}=\int_{t_{1}^{k}}^{t_{2}^{k}} i_{T}^{k}(t)dt$ and $i_{R}^{k}=\int_{t_{1}^{k}-\Delta\tau}^{t_{2}^{k}-\Delta\tau} i_{R}^{k}(t)dt$, where $\Delta\tau$ is the storage time. The current addition for the $k$-th pulse, $i_{+}^{k}= i_{T}^{k}+i_{R}^{k}$, appears when the optical delay is balanced by an electronic delay. Those results are shown in Figs.~\ref{fig:fringes}(b)-(g), in which $i_{T}$ and $i_{R}$ are normalized to keep the splitting ratio as close as possible to $1:1$. The output power of the HD measurements is commonly recorded by a spectrum analyzer, and is proportional to the average $i_{+}^{2}$ over the $N$ repetitions, i.e., $\langle i_{+}^{2}\rangle=(1/N)\sum_{k}^{N}(i_{+}^{k})^{2}$. From the average of the time-domain power, we get the points in Fig.~\ref{fig:fringes}(i), which correspond to the case (I). \par
For the case (II) of the main text, the direct current addition of T and R pulses after the electronic delay compensation is written as $i_{+}^{k}(t)=i_{T}^{k}(t)+i_{R}^{k}(t)$, where $i_{T}(t)$ and $i_{R}(t)$ have been normalized as above. For the integration of the $k$-th pulse from $t_{1}^{k}$ to $t_{2}^{k}$, we write $(i_{+}^{k})^{2}=\int_{t_{1}^{k}}^{t_{2}^{k}}\left[i_{+}^{k}(t)\right]^{2}dt$ to get the value of each individual measurement, and compare it with the result of case (I). Finally, the interference pattern is recovered by the average of the total power of the $N$ repetitions, which is proportional to $\langle i_{+}^{2}\rangle=(1/N)\sum_{k}^{N}(i_{+}^{k})^{2}$, as shown in Fig.~\ref{fig:visibility}(c). \par

\section{Quantum Noise Analysis}

\subsection{Heterodyne detection of time-separated signals}

In our protocol, the transmitted ($T$) and retrieved ($R$) optical pulses are measured independently through local heterodyne detection. The measurement quadratures can be decomposed into mean values and fluctuations, written as 
$X_{j}=\bar{X}_{j}+\delta X_{j}$, for $j=T,R$. Thus, the noise attributes to the fluctuation of the measured photocurrent for each pulses are 
\begin{equation}
    \begin{aligned}
  \delta i_{j}(t)\propto|\mathcal{E}|\int d\tau  k(t-\tau)\delta X_{j}(\tau). 
    \end{aligned}
\end{equation}
The fluctuation of current addition is
\begin{equation}
    \begin{aligned}
  \delta i_{+}(t)\propto|\mathcal{E}|\int d\tau  k(t-\tau)[\delta X_{T}(\tau)+\delta X_{R}(\tau)]. 
    \end{aligned}
\end{equation}
This gives the fluctuation of output power 
\begin{equation}
\begin{aligned}
    &\langle\delta i^{2}_{+}(t)\rangle\propto|\mathcal{E}|^{2}\int d\tau d\tau^{\prime}k(t-\tau)k(t-\tau^{\prime})\\
    &\cdot[\mathcal{C}_{TT}(\tau,\tau^{\prime})+\mathcal{C}_{RR}(\tau,\tau^{\prime)}+\mathcal{C}_{TR}(\tau,\tau^{\prime})+\mathcal{C}_{RT}(\tau,\tau^{\prime})], \\
\end{aligned}
\end{equation}
where $\mathcal{C}_{ij}(\tau,\tau^{\prime})=\langle\delta X_{i}(\tau)\delta X_{j}(\tau^{\prime})\rangle$, $i,j=T,R$ is the fluctuation correlation function. 

Since transmitted and retrieved pulses are measured at different times via independent local heterodyne detections, the vacuum/electronic noise independently introduced by the two detection events is uncorrelated, so that the cross-noise terms $\mathcal{C}_{ij}(\tau,\tau^{\prime})$, for $i\neq j$, vanish and the total current noise becomes the sum of the two individual noises, i.e. $\langle\delta i^{2}_{+}(t)\rangle\approx\langle\delta i^{2}_{T}(t)\rangle+\langle\delta i^{2}_{R}(t)\rangle$. If these two are compatible, then $\langle\delta i^{2}_{+}(t)\rangle\approx2\langle\delta i^{2}_{hom}(t)\rangle$, where $\langle\delta i^{2}_{hom}(t)\rangle$ is the noise that comes from local homodyne detections. Furthermore, for a conventional phase-insensitive heterodyne measurement in which both quadratures are reconstructed, an additional 3-dB quantum-noise penalty arises from the image-band vacuum relative to an ideal single-quadrature homodyne measurement, gives $\langle\delta i^{2}_{het}(t)\rangle=2\langle\delta i^{2}_{hom}(t)\rangle$~\cite{Yuen1983Noise,Buonanno2003Quantum}. \par

\subsection{Phase-estimation noise and magnetic sensitivity}

The analysis above establishes the quantum noise introduced at the optical readout stage. To connect this photocurrent noise to the experimentally relevant magnetic-field sensitivity, we next propagate the current fluctuations through the estimator which was used to reconstruct the relative phase between the transmitted and retrieved temporal modes. 

More generally, denoting the experimentally reconstructed observable by $\hat S$, the corresponding phase uncertainty can be written through error propagation as
\begin{equation}
    \begin{aligned}
    \sigma_{\Delta\phi}^{2}=\frac{\mathrm{Var}(\hat S)}{\left|\partial\langle\hat S\rangle/\partial\Delta\phi\right|^{2}}.
    \end{aligned}
\end{equation}
For the estimator used in the experiment, $\hat S=\langle i_+^2\rangle$, the photocurrent variance derived above therefore directly determines the quantum-limited contribution to $\sigma_{\Delta\phi}$. Since the atomic phase shift induced by the magnetic field is
\begin{equation}
    \begin{aligned}
    \Delta\phi=\gamma B\Delta\tau,
   \end{aligned}
\end{equation}
the corresponding single-shot magnetic field uncertainty is
\begin{equation}
    \begin{aligned}
    \sigma_B=\frac{\sigma_{\Delta\phi}}{\gamma\Delta\tau}.
   \end{aligned}
\end{equation}
For a pulsed experiment with cycle time $T_{\rm cyc}$, the sensitivity normalized to unit bandwidth is consequently
\begin{equation}
    \begin{aligned}
    \eta_B=\frac{\sigma_{\Delta\phi}}{\gamma\Delta\tau}\sqrt{T_{\rm cyc}}.
   \end{aligned}
\end{equation}
Accounting for the visibility degradation due to the decoherence governed by the effective transverse coherence time $T_2^*$, the resulting phase and magnetic-field sensitivity are 
\begin{equation}
    \sigma_{B}=\frac{\sigma_{\Delta\phi}}{\mathcal{V}^*\gamma \Delta\tau},\quad \eta_{B}=\frac{\sigma_{\Delta\phi}}{\mathcal{V}^*\gamma \Delta\tau}\sqrt{T_{\rm cyc}},
\end{equation}
where $\mathcal{V}^{*}=\mathcal{V}e^{-\Delta\tau/T_2^*}$. 

The phase uncertainty measured experimentally contains not only the fundamental optical readout noise discussed above, but also atomic projection noise and technical fluctuations. It is therefore useful to separate the total variance into quantum and classical contributions,
\begin{equation}
    \begin{aligned}
    \sigma_{\Delta\phi}^{2}=\sigma_Q^{2}+\sigma_{\rm cl}^{2},
   \end{aligned}
\end{equation}
where
\begin{equation}
    \begin{aligned}
    \sigma_Q^{2}=\sigma_{\rm photon}^{2}+\sigma_{\rm atom}^{2},
   \end{aligned}
\end{equation}
and
\begin{equation}
\sigma_{\rm cl}^{2}=\sigma_{\rm electric}^{2}+\sigma_{\rm path}^{2}+\sigma_{\rm LO}^{2}+\sigma_{\rm resB}^{2}+\cdots.
\end{equation}
A phenomenological effective readout factor may then be defined as
\begin{equation}
    \begin{aligned}
    \eta_{\rm eff}\equiv\frac{\sigma_Q}{\sigma_{\Delta\phi}},
   \end{aligned}
\end{equation}
such that
\begin{equation}
    \begin{aligned}
    \sigma_{\Delta\phi}^{2}=\frac{\sigma_Q^{2}}{\eta_{\rm eff}^{2}}.
   \end{aligned}
\end{equation}
In the present regime, where the optical excitation is much smaller than the number of atoms participating in the collective spin wave, photon shot noise is expected to dominate over atomic projection noise. Under this approximation,
\begin{equation}
    \begin{aligned}
    \sigma_{\Delta\phi}\simeq\frac{1}{\eta_{\rm eff}\sqrt{N_{\rm photon}}},
   \end{aligned}
\end{equation}
so that excess technical noise may equivalently be interpreted as reducing the effective photon number available for phase estimation to
\begin{equation}
N_{\rm eff}=\eta_{\rm eff}^{2}N_{\rm photon}.
\end{equation} \par

\subsection{Comparison with direct optical interference}

Having expressed both the optical readout noise and the additional experimental noise sources in terms of the common phase uncertainty $\sigma_{\Delta\phi}$, we can now compare the present electronic-reconstruction scheme with direct optical interference on equal footing. Both approaches estimate the same accumulated atomic phase $\Delta\phi=\gamma B\Delta\tau$; they differ only in how the two temporally separated optical modes are brought into a common measurement basis. 

In direct optical interference, one temporal mode must physically propagate through an optical delay line before interfering with the other. A lossy delay channel with power transmission $\mathcal T(l)$ can be represented as
\begin{equation}
\hat a_{\rm out}=\sqrt{\mathcal T(l)}\,\hat a_{\rm in}+\sqrt{1-\mathcal T(l)}\,\hat v,
\end{equation}
where $\hat v$ is the vacuum mode introduced by loss. For an attenuation coefficient $\alpha$, $\mathcal T(l)=e^{-\alpha l}$, or equivalently $\mathcal T(l)=10^{-a l/10}$ when the fiber attenuation $a$ is expressed in dB per unit length. For a coherent-state input, the coherent field amplitude and hence the interferometric signal slope decrease as $\sqrt{\mathcal T(l)}$, whereas the normalized vacuum quadrature noise remains at the shot-noise level. The resulting phase and magnetic-field sensitivities therefore deteriorate approximately as
\begin{equation}
\sigma_{\Delta\phi}^{(\mathrm{delay})}
\propto
\mathcal T(l)^{-1/2},
\qquad
\eta_B^{(\mathrm{delay})}
\propto
\mathcal T(l)^{-1/2}.
\end{equation}

\begin{figure}[htbp]
    \centering
    \includegraphics[width=1.0\linewidth]{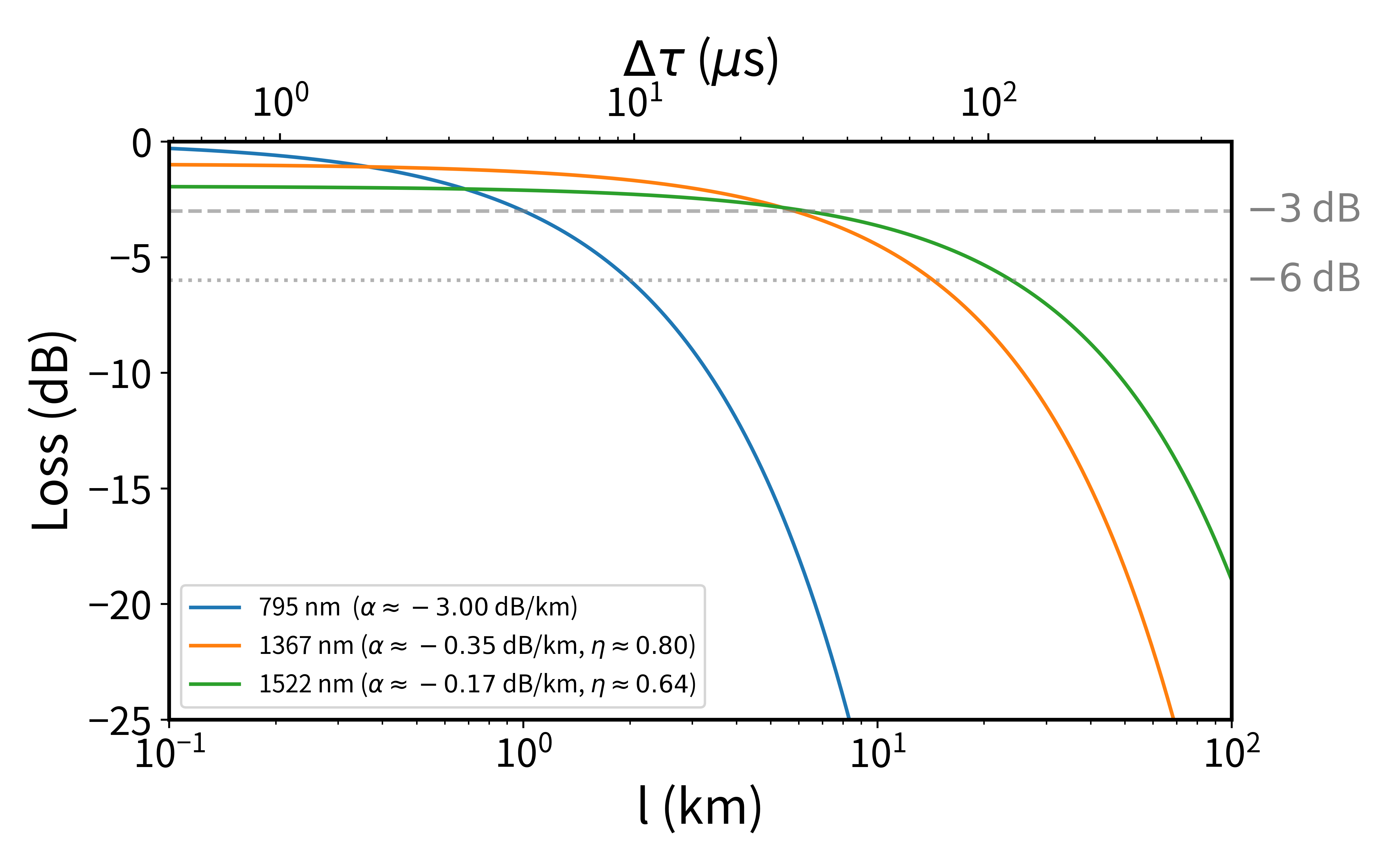}
    \caption{Comparing the noise level of homodyne detection, \SI{-3}{dB} and heterodyne detection, \SI{-6}{dB} and fiber losses for \SI{795}{nm}, \SI{1367}{nm} and \SI{1522}{nm}. The \SI{1367}{nm} and \SI{1522}{nm} are frequency converted from \SI{795}{nm}, with conversion efficiency $\eta=0.80$~\cite{Chen2025Highefficiency} and $\eta=0.64$\footnote{This conversion efficiency is actually from \SI{780}{nm} to \SI{1522}{nm}.}~\cite{Luo2025Entangling}. }
    \label{fig:fiberloss}
\end{figure}

The two architectures thus exhibit qualitatively different quantum-noise penalties. Direct optical interference preserves a single coherent optical comparison but becomes increasingly affected by propagation loss and the associated vacuum fluctuations as the required delay increases. For long time-delays requiring kilometer-scale optical fibers, the propagation-induced attenuation becomes a major limitation, as shown in Fig.~\ref{fig:fiberloss}. Even we consider a conversion to the low loss communication band frequency, including the conversion efficiency, the transmission loss will lead an inferior comparing with homodyne and heterodyne detection when delay lines go to dozens of kilometers. In contrast, the present electronic reconstruction avoids long-distance optical propagation and its associated loss, at the price of performing two local optical measurements whose measurement-added noises contribute to the reconstructed phase difference. This trade-off provides a common framework for comparing the two approaches as the temporal separation, and hence the equivalent optical-delay length, is increased. \par

\end{appendix}

\end{document}